\documentclass[twocolumn,preprintnumbers,superscriptaddress,nofootinbib,aps,prd,floatfix]{revtex4}

\pdfoutput=1

\usepackage{amsmath,amssymb}
\usepackage{graphicx} 
\usepackage{epstopdf} 
\usepackage{slashed}
\usepackage{color}
\usepackage{multirow}
\usepackage{footmisc}
\usepackage{hyperref}
\usepackage{subfigure}
\usepackage{snapshot}
\usepackage{hepunits}
\usepackage{tabularx}

\usepackage{array}
\newcolumntype{L}[1]{>{\raggedright\let\newline\\\arraybackslash\hspace{0pt}}m{#1}}
\newcolumntype{C}[1]{>{\centering\let\newline\\\arraybackslash\hspace{0pt}}m{#1}}
\newcolumntype{R}[1]{>{\raggedleft\let\newline\\\arraybackslash\hspace{0pt}}m{#1}}

\DefineFNsymbols*{lamport}{\dagger\ddagger\S\P\|{**}{\dagger\dagger}{\ddagger\ddagger}}

\renewcommand{\thefootnote}{\fnsymbol{footnote}}

\renewcommand{\d}{{\mathrm{d}}}
\newcommand{\be}{\begin{eqnarray*}}
\newcommand{\ee}{\end{eqnarray*}}

\newcommand{\bee}{\begin{eqnarray}}
\newcommand{\eee}{\end{eqnarray}}
\newcommand{\beeq}{\begin{equation}}
\newcommand{\eeeq}{\end{equation}}

\newcommand{\fb}{{\text{fb}}}

\renewcommand{\vec}{\bf}

\begin{document}

\title{Uncovering the relation of a scalar resonance to the
  Higgs boson}
%
%

\begin{abstract}
We consider the associated production of a scalar resonance with the standard model Higgs boson. We
demonstrate via a realistic phenomenological analysis that couplings of such a resonance to the Higgs boson can be constrained
in a meaningful way in future runs of the LHC, providing insights on its origin and its
relation to the electroweak symmetry breaking sector. Moreover, the final state
can provide a direct way to determine whether the new resonance is
produced predominantly in gluon fusion or quark-anti-quark annihilation. The analysis focusses on a resonance coming from a scalar field with vanishing vacuum expectation value and its decay to a
photon pair. It can however be straightforwardly generalised to other scenarios.

\end{abstract}

\author{Adri\'an Carmona}
\email{adrian.carmona@cern.ch}
\affiliation{CERN, Theoretical Physics Department,\\ CH-1211 Geneva 23, Switzerland}

\author{Florian Goertz}
\email{florian.goertz@cern.ch}
\affiliation{CERN, Theoretical Physics Department,\\ CH-1211 Geneva 23, Switzerland}

\author{Andreas Papaefstathiou} 
\email{apapaefs@cern.ch}
\affiliation{CERN, Theoretical Physics Department,\\ CH-1211 Geneva 23, Switzerland}

\preprint{CERN-TH-2016-133, MCnet-16-19}

\maketitle

\section{New scalar resonances at colliders.}
\label{sec:intro}

Models with an additional (pseudo-)scalar singlet with a mass of several hundred GeV represent a well motivated class of extensions of the Standard Model (SM) of particle physics, including composite Higgs scenarios, supersymmetry, Coleman-Weinberg models, models addressing the strong CP problem, models of flavor, as well as generic Higgs portal setups (see e.g. \cite{Falkowski:2015iwa,Bellazzini:2014yua,Bellazzini:2015nxw,Ellwanger:2009dp,Hempfling:1996ht,Englert:2013gz,Gherghetta:2016fhp,Tsumura:2009yf,Bauer:2016rxs}). A particularly promising channel to search for and analyze such a particle is its decay to two photons. Beyond being possibly sizable in certain scenarios, it offers a robust and clean way to detect a signal, emerging over a steeply falling background \cite{Khachatryan:2014ira,Aad:2014eha}.

After its discovery, an important aspect of scrutinising any new resonance is in fact to
measure its couplings, and hence determine its relation, to the known particle content of the Standard
Model (SM). A crucial component of this task is to uncover its role
in the arena of electroweak symmetry breaking
(EWSB). As a first step in this direction, determination of the couplings of the new scalar to
the SM-like Higgs boson is mandatory, which is the main focus of this article,
employing its di-photon ($\gamma\gamma$) decay channel.\footnote{A specific motivation for the first version of this manuscript was provided by the
apparent $\gamma\gamma$ resonance at $M_{\gamma\gamma} \sim 750$\,GeV in
ATLAS~\cite{ATLAS-CONF-NOTE, ATLAS-CONF-2016-018} and CMS~\cite{CMS:2015dxe, CMS-PAS-EXO-16-018} data. This turned out not to be present in the 2016 data \cite{ATLAS:2016eeo,Khachatryan:2016yec}. Consequently the article was generalized to other mass scales of a potential scalar resonance, which remains well motivated, taking into acount new limits on its cross section - see below. Comprehensive analyses studying constraints on (other) possible couplings of a di-photon resonance as well as detailed examinations of indirect footprints of new (high multiplicity) sectors, linked to its productions or decay appeared {\it e.g.} in
\cite{Franceschini:2015kwy,Angelescu:2015uiz,Knapen:2015dap,DiChiara:2015vdm,McDermott:2015sck,Ellis:2015oso,Gupta:2015zzs,Falkowski:2015swt,Alves:2015jgx,Goertz:2015nkp,Son:2015vfl,Gao:2015igz,Salvio:2015jgu,
Gu:2015lxj,Djouadi:2016eyy,Salvio:2016hnf,Gross:2016ioi,Goertz:2016iwa,Bernon:2016dow,Panico:2016ary,Kamenik:2016tuv,Chala:2016mdz,Franceschini:2016gxv,Buttazzo:2015txu,Berthier:2015vbb,Dev:2015vjd,Roig:2016tda,Dawson:2016ugw}.}

For a $\gamma\gamma$ resonance originating from a scalar field $\mathcal{S}$, neutral under the
SM gauge group, the relevant effective Lagrangian for our study -- augmenting the SM at dimension $D\leq 5$ -- is
\bee\label{eq:lagrangian}
\mathcal{L}_{\rm eff}
&\supset& \frac{1}{2}\partial_\mu {\cal S} \partial^\mu {\cal S} - \frac 1 2 \mu_S^2 {\cal S}^2 \\
&-&(y_d^S)^{ij} \frac{\cal{S}}{\Lambda}\bar{Q}_L^{i}H d_R^{j} - (y_u^S)^{ij}\frac{\cal{S}}{\Lambda}\bar{Q}_L^i\tilde{H}u_R^j +\mathrm{h.c.}\nonumber\\
&-&\frac{\cal{S}}{\Lambda}\frac{1}{16\pi^2}\left[g^{\prime 2} c_B^S B_{\mu\nu} B^{\mu\nu}+ g^2 c_W^S W^{I\mu\nu} W_{\mu\nu}^I\right.\nonumber\\
&+&\left.g_S^2 c_G^S G^{a\mu\nu}G_{\mu\nu}^a\right]- \lambda_{HS} |H|^2 {\cal{S}}^2 - \frac{\lambda_{S}}{4} {\cal{S}}^4\,.\nonumber
\eee
Here, $Q_L^i$ is the $i$-th generation left-handed $SU(2)_L$ fermion
doublet, $d^j_R$ and $u^j_R$ are the right-handed $SU(2)_L$ fermion singlets for
generation $j$, $(y^S_q)^{ij}$ are the corresponding Yukawa-like 
couplings, $c_B^S$, $c_W^S$ are the couplings of $\mathcal{S}$ to the $U(1)_Y$ and $SU(2)_L$ 
gauge fields $B$ and $W$, $c_G^S$ is the coupling to the gluon
fields, $H$ is the Higgs boson doublet, $\lambda_{HS}$ is the Higgs-Scalar portal
coupling and $\lambda_{S}$ the new scalar quartic. Moreover, $\Lambda$ denotes
the scale of heavy new physics (NP), mediating the contact interactions of $\mathcal{S}$
with SM gauge bosons and fermions (the latter involving $H$ to generate a 
gauge singlet). 
Note that we do not include terms with an odd number
of $\cal{S}$ fields containing only scalars (as well as lepton fields). The corresponding interaction vertices will turn out 
irrelevant in general for the process we will consider, see below.\footnote{The full list of potential $D\leq 5$ operators
is listed in Appendix~\ref{app:operators}.}  Beyond that, terms linear in $\cal{S}$ could also lead to the singlet mixing with the Higgs boson after EWSB, which would in fact affect its phenomenology.
Although such effects could still be present at a non-negligible level, they are expected to be sub-leading and we neglect them for simplicity,
see Appendix~\ref{app:fullfit}.
Furthermore, the analysis that follows is independent of the CP properties of $\mathcal{S}$ and its interactions
and henceforth, for simplicity, we assume it to be CP-even with CP-conserving interactions.

With the potential for the Higgs doublet $H$ taking the conventional form:
\bee 
\mathcal{V} = \lambda_H |H|^4 - \mu^2_H |H|^2\,,
\eee
and assuming that $H$ is the only scalar that gets a vacuum expectation value (vev), $|\langle H \rangle |= v/\sqrt 2$, triggering EWSB, we obtain the condition\footnote{The fact that 
$\langle {\cal S} \rangle$=0 guarantees the full absence of scalar mixing, that could otherwise occur even
without linear terms in $S$.} 
\bee
\lambda_{HS} \mu_H^2 - \lambda_H \mu_S^2 < 0\,,\quad \mu_H^2 < 0\,,
\eee
The physical mass of the singlet thus reads $M = \sqrt{\mu_S^2 + \lambda_{HS} v^2}\,$.

The resulting trilinear interactions between the physical scalar resonances after EWSB 
are described by
\begin{equation}\label{eq:indep}
\mathcal{L}_\mathrm{scalar}^3 =  - \frac{ M_h^2 } { 2 v } h^3 
- \lambda_{HS}\, v\, h S^2 \;,
\end{equation}
where $h$ is the Higgs boson, which (due to the case of negligible
scalar mixing) is basically fully embedded in $H$ and describes excitations around its vev, 
such that in unitary gauge 
$H \simeq 1/\sqrt2 (0, v + h)^T$, and $S$ is a new scalar resonance, which can also be (approximately)
identified as ${\cal S}=S$. Moreover, $v
\simeq 246$~GeV is the Higgs vev, $M_h \simeq
125$~GeV is the measured Higgs boson mass and  
the portal coupling $\lambda_{HS}$ is to be determined, being a main scope of this
paper. This coupling would be basically unconstrained by direct observation of a di-photon resonance.
Nevertheless, loose indirect constraints can be derived, 
requiring vacuum stability not to be spoiled.
They read
\beeq
\lambda_S>0\,,\quad \lambda_H>0\,,\quad \lambda_{HS}^2<\lambda_H\lambda_S\,,
\eeeq
and need to be imposed at least at the scale where the NP enters, {\it i.e.}, 
the TeV scale (see, {\it e.g.} \cite{Falkowski:2015iwa,Zhang:2015uuo}).\footnote{While 
the first two conditions need to hold at all scales, for $\lambda_{HS}>0$ 
the last condition might be violated at higher scales, while still the electroweak 
vacuum remains stable \cite{Falkowski:2015iwa}. Moreover, for the latter condition 
odd terms in ${\cal S}$ are assumed to vanish.}
Requiring $\lambda_H \sim 0.13$, to fit the observed Higgs mass, as well 
as $\lambda_S < (4 \pi)^2$, we thus obtain
$-4.5 \lesssim \lambda_{HS} \lesssim 4.5$. We will see below that in general 
our analysis can put stronger bounds than these on $\lambda_{HS}$.\footnote{
Note that requiring a more conservative limit, such as $\lambda_S < {\cal O}(10)$ (corresponding to, {\it e.g.}, $d\lambda_S/\lambda_S < 1$ \cite{Falkowski:2015iwa}), restricts $\lambda_{HS}$ to be not much larger than 1 and would remove a considerable portion of the parameter space where our analysis exhibits sensitivity. However, in any case, the limits presented here are complementary to such considerations.}

In the present study we consider measurement of the coupling 
$\lambda_{HS}$ at the LHC, where it can be probed via associated
production of the new resonance with the SM-like Higgs boson: $pp
\rightarrow hS$. 
For this process, the interactions neglected in
(\ref{eq:lagrangian}) play no role to good approximation:
they would either not enter at leading order (LO), or, as is the case for the $|H|^{2,4} {\cal S}$ interactions,
contribute at most to a diagram with a (strongly suppressed) off-shell Higgs boson propagator, for details see the appendices~\ref{app:operators} and~\ref{app:fullfit}.

In principle several decay modes of $S$ can be
considered. Here we focus on the process $pp \rightarrow hS
\rightarrow h \gamma\gamma $, where the new particle decays to a pair
of photons. Given that the tentative cross section of the resonant di-photon production,  $pp \rightarrow S
\rightarrow \gamma\gamma$, will be known and will be well-measured in the case of discovery, this allows constraints on the coupling $\lambda_{HS}$ to be
imposed almost independently of the couplings to the initial-state
partons and final-state photons, given that only a single production mode is relevant. We consider production via
gluon fusion and quark-anti-quark annihilation, mediated through non-vanishing
coefficients $c_G^S$, $(y_{d}^{S})^{22}$ or $(y_{d}^{S})^{33}$, respectively, and show how these modes
could be disentangled via appropriate measurements. We will implicitly assume not too large values of $\Gamma(S\to \gamma\gamma)$, in such a way that photo-production is always subdominant.

Although we will focus on three specific benchmark masses of $M=600,~750,~900$~GeV, our analysis could be applied to the general case of associated production of a Higgs boson with a scalar di-photon resonance of any mass. Moreover, several
features of the final state studied here, such as the invariant mass of
the final-state scalar or the total invariant mass of the process, will exhibit similar features
when considering other decay modes.

The article is organised as follows: in Section~\ref{sec:hS} we
examine the process of associated production of a scalar resonance and
a Higgs boson, in Section~\ref{sec:sim} we describe the event
generation and detector simulation setup and in
Section~\ref{sec:analysis} we provide details of the analysis and
results. Finally, we conclude in Section~\ref{sec:conc}. 

\section{Associated production of a scalar resonance and a Higgs boson.}
\label{sec:hS}

\subsection{Production through gluon fusion.}
\begin{figure}[!h]
 \centering
  \subfigure[]{\includegraphics[width=0.26\textwidth]{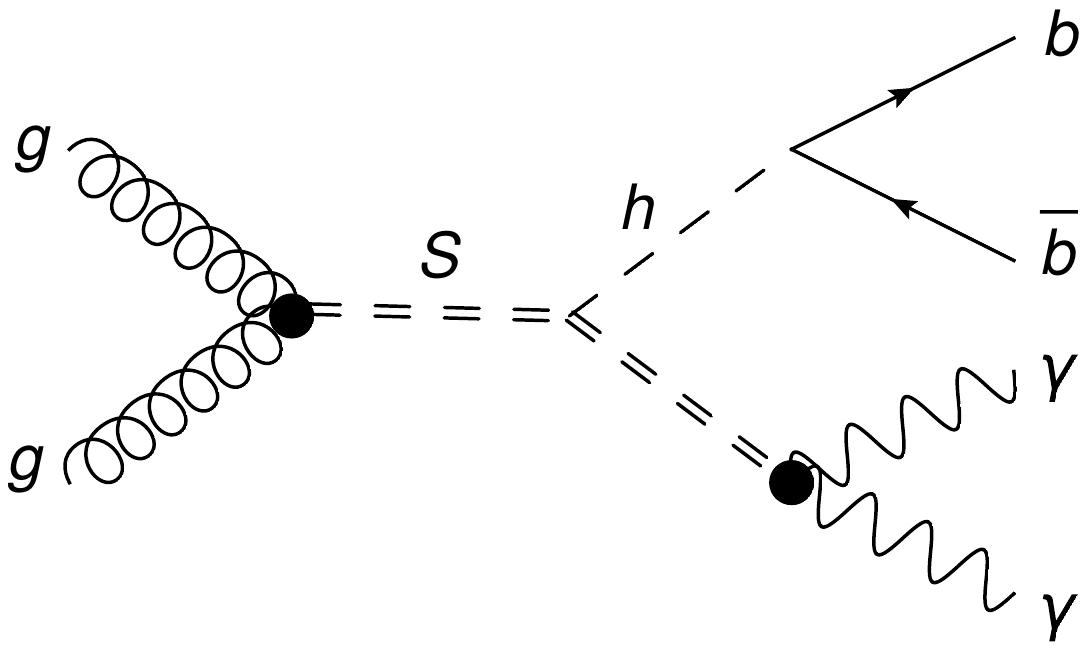}}
   \hfill
   \subfigure[]{\includegraphics[width=0.23\textwidth]{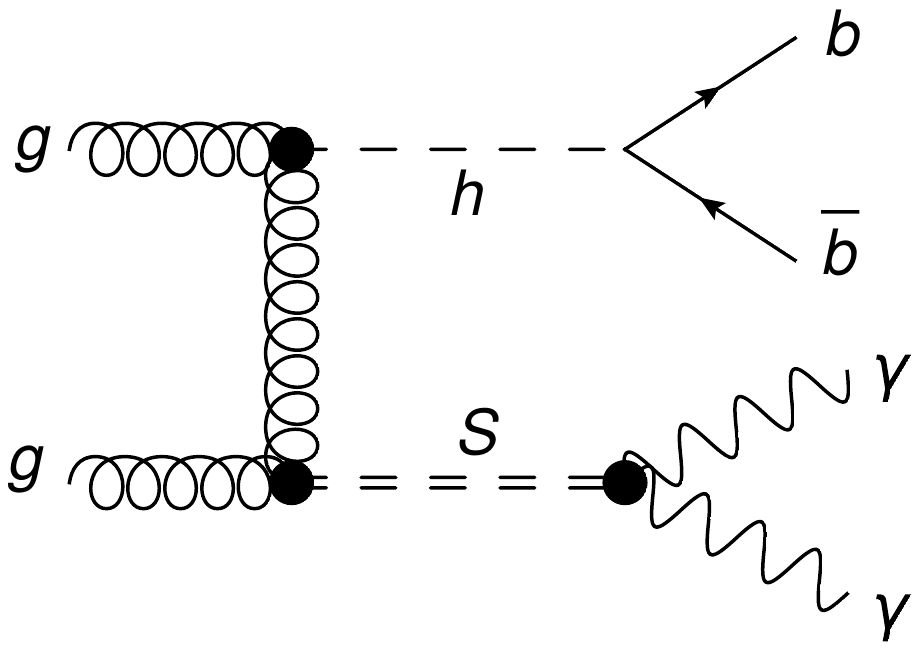}}
  \caption{\label{fig:ggdiags} The diagrams contributing to the
    process $gg \rightarrow h S \rightarrow (b\bar{b}) (\gamma
    \gamma)$ at the LHC at LO.}
\end{figure}

The dominant diagrams at LO contributing to the production of the $hS$ final state in gluon fusion
and subsequent decay of the resonance $S$ to a pair of photons via the interactions
\begin{align}
	\mathcal{L}_{\rm eff}&\supset -\frac{S}{\Lambda}\frac{1}{16\pi^2}\left[ e^2(c_B^S+c_W^S)F_{\mu\nu}F^{\mu\nu}  + g_s c_G G^{a\mu\nu} G^a_{\mu\nu}  \right]\nonumber \\ &\, =-\frac{S}{\Lambda}\frac{1}{16\pi^2} \left[ e^2 c_{\gamma}^SF_{\mu\nu}F^{\mu\nu} + g_s c_G G^{a\mu\nu} G^a_{\mu\nu} \right] \,,
\end{align}
are shown in Fig.~\ref{fig:ggdiags}. 
In the analysis of the present article, we will consider the Higgs boson decaying to a bottom-quark pair,
since this maximizes the expected number of events, which would be modest in general. There exist both the $s$-channel $S$ exchange, involving the portal coupling $\lambda_{HS}$ and depicted
in the upper panel (a), as well as the `direct' $hS$ production, via $t$-channel gluon exchange, depicted in the lower panel (b).

\subsection{Production through quark-anti-quark annihilation.}

\begin{figure}[!h]
 \centering
  \subfigure[]{\includegraphics[width=0.26\textwidth]{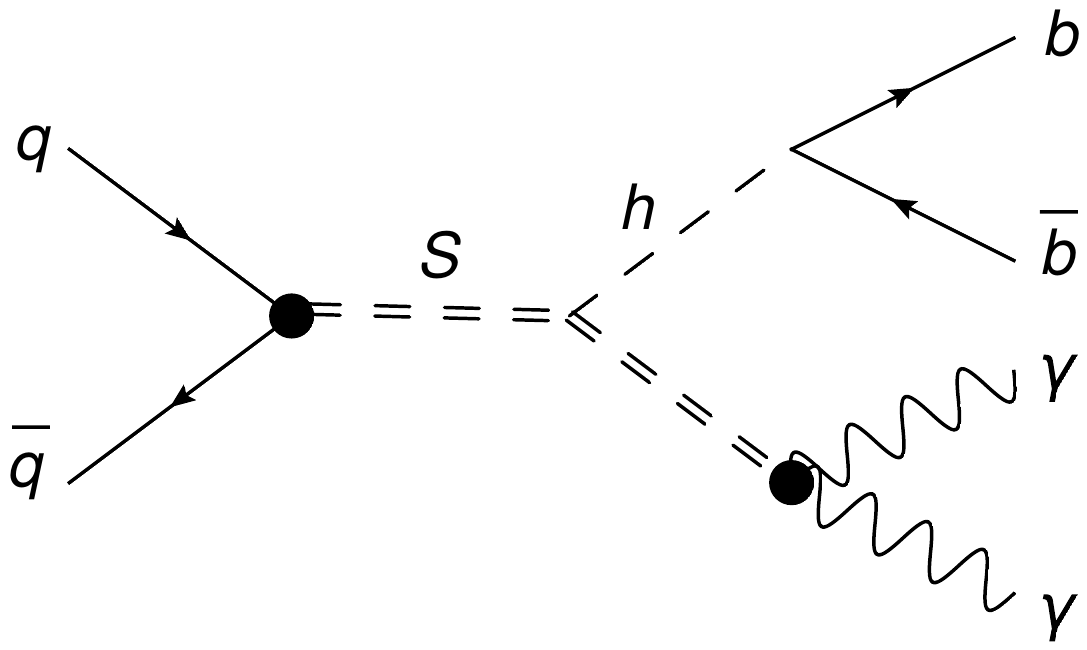}}
   \hfill
   \subfigure[]{\includegraphics[width=0.23\textwidth]{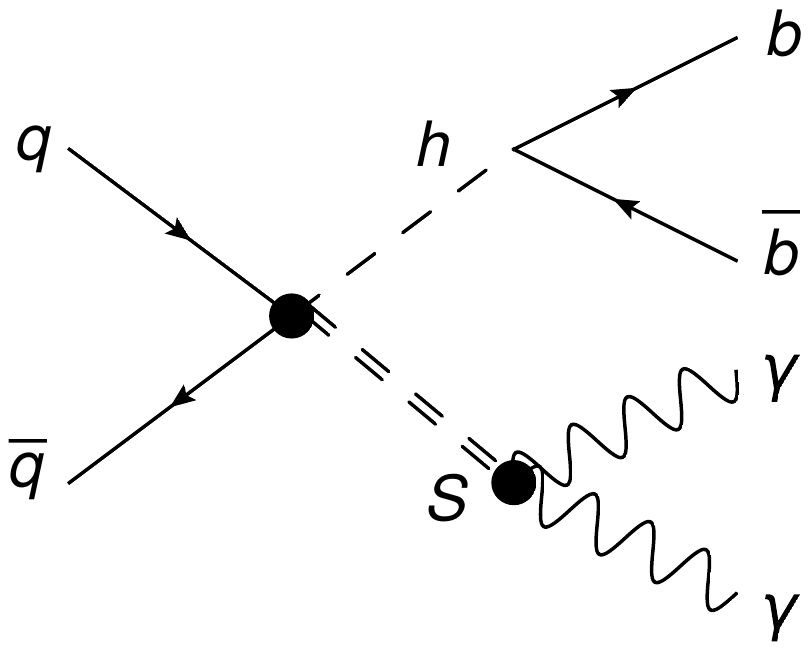}}
  \caption{\label{fig:qqdiags} The diagrams contributing to the
    process $q\bar{q} \rightarrow h S \rightarrow (b\bar{b}) (\gamma
    \gamma)$ at the LHC at LO.}
\end{figure}

For the case of quark anti-quark annihilation, a new
diagram arises from the contact interaction $q\bar{q}hS$.\footnote{The
  t-channel diagram with the $q\bar{q}h$ interaction is suppressed due
  to a small Yukawa coupling.} Both
contributing graphs are shown in Fig.~\ref{fig:qqdiags}. The new
diagram (b) distinguishes the $q\bar{q}$ annihilation from the gluon
fusion case. An important fact is that now the $hS$ process is
non-vanishing and significant even in the absence of the portal coupling
$\lambda_{HS}$. This indicates that one can employ this final state to
exclude $q\bar{q}$ annihilation as the dominant production process
(in the absence of a signal). In the following we will focus on the cases of $q=b$, $q=s$ or $q=c$.

It will be useful in both scenarios to construct the ratio of the
associated production process $pp \rightarrow hS$, through all possible intermediate
states, with the subsequent decay of the
new resonance to a $\gamma\gamma$ final state, to that of the single production $pp \rightarrow S \rightarrow
\gamma \gamma$,
\bee
\label{eq:rho}
\rho(xx') = \frac{ \sigma(xx'\rightarrow hS \rightarrow h \gamma
  \gamma) } {  \sigma(xx'\rightarrow S 
\rightarrow \gamma \gamma) }\,,
\eee
where we consider $xx' = \{gg, b\bar{b}, s\bar{s}\,, c\bar{c}\,\}$.\footnote{
In the most general setup, the analysis of this article can constrain the
sum of the squares of the couplings of $S$ to all quark
generations (for a given $\lambda_{HS}$), appropriately weighted by the parton density functions.}
The ratio is a useful quantity since it removes the
dependence on the product of couplings of the new resonance to
the initial-state partons and final-state photons. Moreover, it can be
used to absorb, at least approximately, theoretical and experimental
systematic uncertainties.\footnote{For a similar idea investigated in
  the context of Higgs boson pair production,
  see~\cite{Goertz:2013kp}.}

\begin{figure}[!h]
 \centering
  \includegraphics[width=0.42\textwidth]{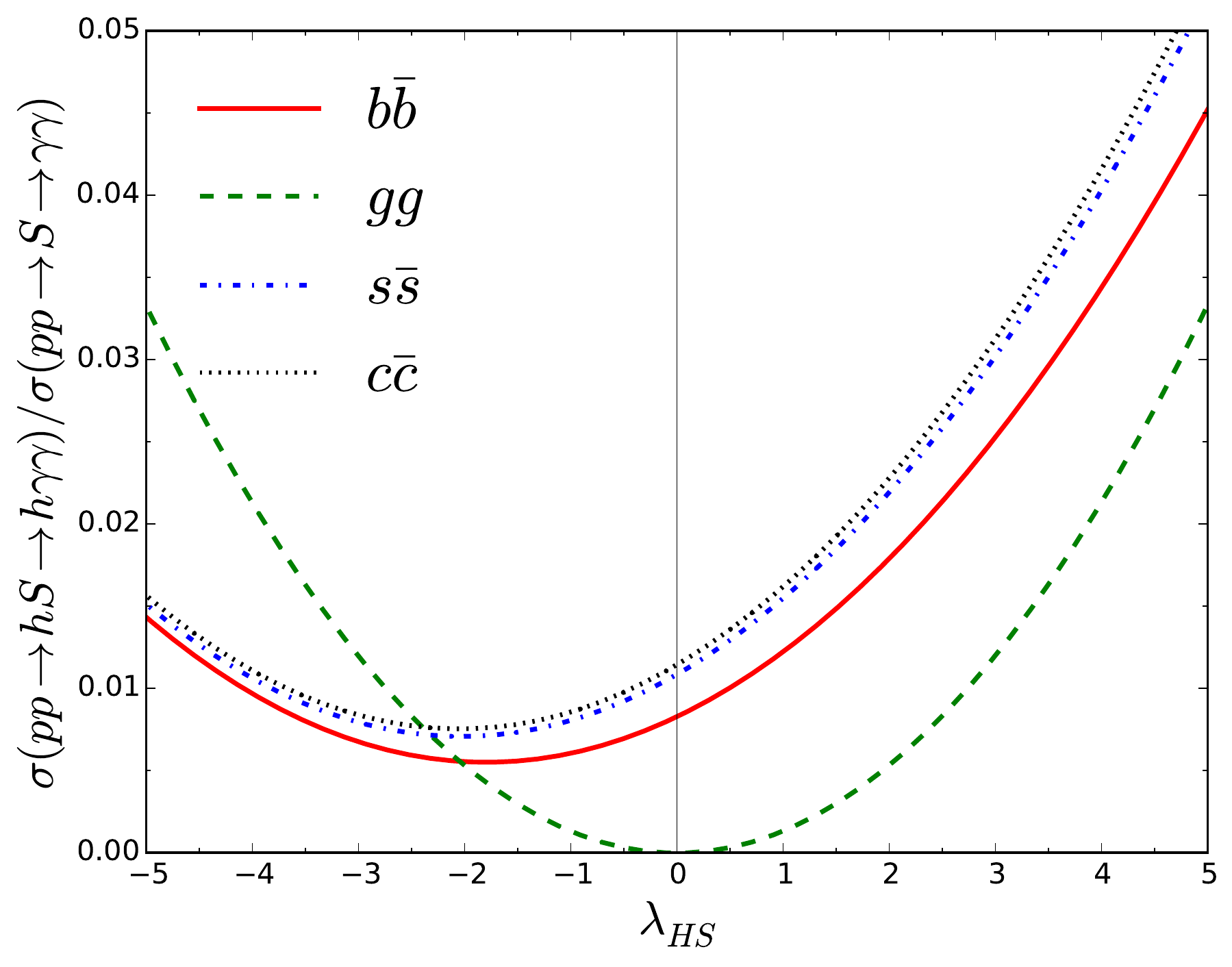}
  \caption{\label{fig:ratiofits} The ratios $\rho(gg)$,
    $\rho(b\bar{b})$, $\rho(s\bar{s})$ and $\rho(c\bar{c})$ for $gg$, $b\bar{b}$,
    $s\bar{s}$ and $c\bar{c}$ initial states respectively, defined between the associated production $pp \rightarrow
    hS \rightarrow h\gamma\gamma$ and the single
    production $pp\rightarrow S \rightarrow \gamma \gamma$, as functions of the portal coupling $\lambda_{HS}$. The mass of the scalar resonance was taken to be $M = 750$~GeV and the width $\Gamma =1$~GeV.}
\end{figure}

We show the dependence of the ratio $\rho$ on the portal coupling
$\lambda_{HS}$ in Fig.~\ref{fig:ratiofits} for $gg$, $b\bar{b}$, $s\bar{s}$ and $c\bar{c}$ initial
states for the example di-photon resonance mass $M = 750$~GeV and width $\Gamma = 1$~GeV.\footnote{We employ a single cut of $M_{\gamma \gamma}>200\,$GeV at generation level in order to remove (SM-like) $pp \to hh \to h \gamma\gamma$ interference with the signal, {\it i.e.}, $S(\gamma\gamma)+h$ production. Only after this cut, we can identify a `signal' contribution to the actual physical process - which is Higgs production in association with a photon pair - unambiguously with the process $pp \to h S \to h \gamma \gamma$ to good approximation, assuming the model (\ref{eq:lagrangian}).} A width of $\Gamma \lesssim 1$~GeV can be obtained for example, if $c_\gamma^S \sim \mathcal{O}(10)$ and $c_G \sim \mathcal{O}(1)$ or $(y_d^S)^{33} \sim \mathcal{O}(1)$, for $\Lambda = 1$~TeV, with a cross section in $pp \rightarrow \gamma \gamma$ compatible with current constraints. Similar behaviour of the ratio $\rho$ is obtained for different scalar $S$ masses and widths.

Since the dominant matrix-element contribution to the $gg$-initiated $hS$
process is proportional to the portal coupling, the process approximately vanishes
as $\lambda_{HS} \rightarrow 0$, and hence $\rho(gg) \simeq \rho_{2gg}
\lambda_{HS}^2$, where $\rho_{2gg} \approx 0.00133$, for $M=750$~GeV and $\Gamma = 1$~GeV, obtained by performing a quadratic fit of the $gg$ curve in Fig.~\ref{fig:ratiofits}. As already discussed, this does not hold for the
$q\bar{q}$-initiated process due to the contact interaction
diagram. This results in a non-negligible minimum for $\rho(q\bar{q})$. A
fit to the cross section, again for $M=750$~GeV and $\Gamma = 1$~GeV, yields $\rho(q\bar{q}) \simeq \rho_{2q\bar{q}}
\lambda_{HS}^2 + \rho_{1q\bar{q}} \lambda_{HS} + \rho_{0q\bar{q}}$
with $\rho_{0b\bar{b}} \approx 0.00828$, $\rho_{1b\bar{b}} \approx
0.00309$, $\rho_{2 b\bar{b}} \approx 0.00086$, corresponding to
$b\bar{b}$ initial states, and $\rho_{0s\bar{s}} \approx 0.01025$,
$\rho_{1s\bar{s}} \approx 0.00371$, $\rho_{2 s\bar{s}} \approx 0.00091$, corresponding to
$s\bar{s}$ initial states. The case of $c\bar{c}$ is similar to the $s\bar{s}$ case and therefore in the rest of the article we focus on the cases $q=b$ and $q=s$. The positivity of
the coefficient $\rho_{1q\bar{q}}$ indicates constructive interference
  between the contact interaction and resonant diagrams (for $\lambda_{ HS}>0$). For an extended fit of the ratio $\rho$, including additional diagrams with the production of an intermediate
  Higgs boson due to ${\cal S}|H|^{2,4}$ interactions, that turn out to be sub-dominant, see Appendix~\ref{app:fullfit}. 

The fact that for quark-anti-quark annihilation the $hS$ process is
non-vanishing for all values of the portal coupling
$\lambda_{HS}$ indicates that one could employ this final state to
exclude $b\bar{b}$, $s\bar{s}$ or $c\bar{c}$ annihilation as the dominant production
process. The analysis that will follow in the present article suggests however, that the di-photon decay of the $S$ alone may not be sufficient for that purpose for the benchmark points that we consider. 

\begin{figure}[!h]
 \centering
  \includegraphics[width=0.42\textwidth]{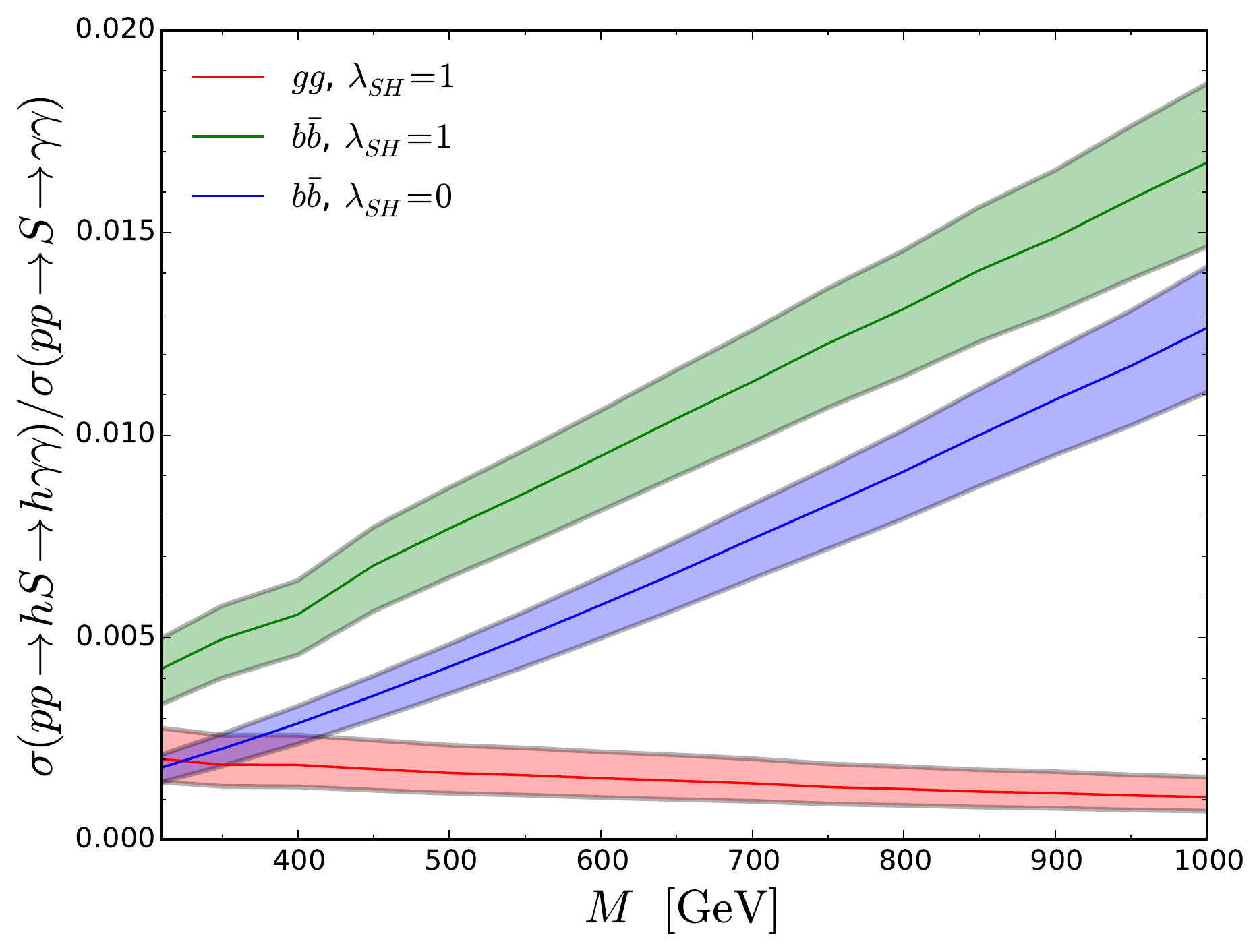}
  \caption{\label{fig:ratiowidthmass} The ratios $\rho(gg)$ and $\rho(b\bar{b})$ for $gg$ and $b\bar{b}$ initial states, defined between the associated production $pp \rightarrow
    hS \rightarrow h\gamma\gamma$ and the single
    production $pp\rightarrow S \rightarrow \gamma \gamma$, as functions of the mass of the resonance, for $\Gamma
    = 1$~GeV.  The bands display the parton density function uncertainty for the
\texttt{MMHT14nlo68cl} set combined in
quadrature with the scale variation between 0.5 and 2.0 times the
default central dynamical scale implemented in
\texttt{MadGraph5\_aMC@NLO}.}
\end{figure}

\begin{figure}[!h]
 \centering
  \includegraphics[width=0.42\textwidth]{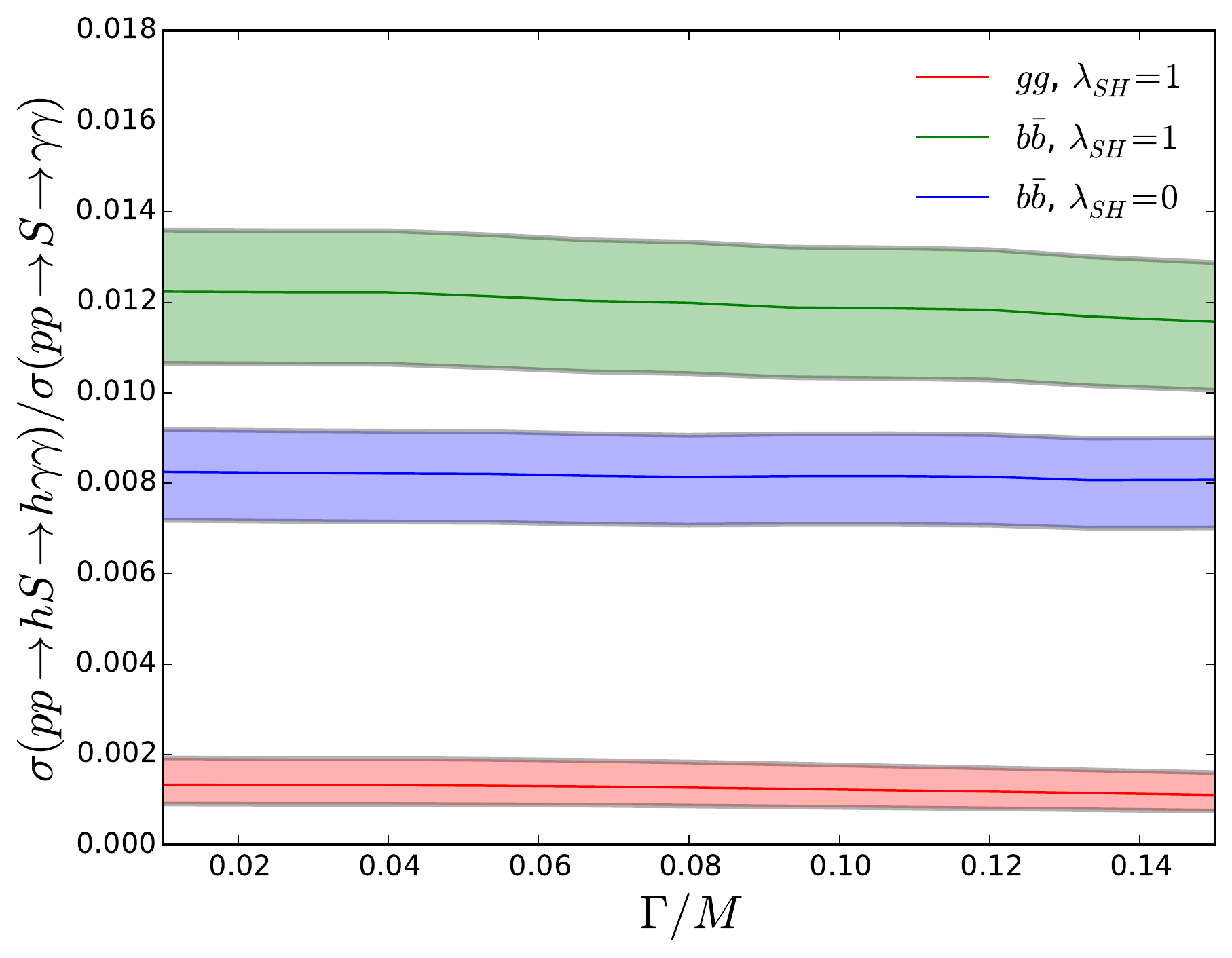}
  \caption{\label{fig:ratiowidth} The ratios $\rho(gg)$ and $\rho(b\bar{b})$ for $gg$ and $b\bar{b}$ initial states, defined between the associated production $pp \rightarrow
    hS \rightarrow h\gamma\gamma$ and the single
    production $pp\rightarrow S \rightarrow \gamma \gamma$, as functions of the width of the resonance over the mass, $\Gamma
    / M$.  The bands display the parton density function uncertainty for the
\texttt{MMHT14nlo68cl} set combined in
quadrature with the scale variation between 0.5 and 2.0 times the
default central dynamical scale implemented in
\texttt{MadGraph5\_aMC@NLO}. The scalar resonance mass was chosen to be $M=750$~GeV.}
\end{figure}

We show in Fig.~\ref{fig:ratiowidthmass} the variation of the ratio $\rho$ with the mass of the resonance, $M$, for the $gg$-iniated process and $\lambda_{HS} =1$, and for the $b\bar{b}$-initiated process for $\lambda_{HS} = 0$ (no portal) and $\lambda_{HS} = 1$. We have fixed the width to $\Gamma = 1$~GeV. Interestingly, the pure $q\bar{q}$-induced processes exhibit an increase of the ratio $\rho$ with increasing mass -- related to the new $q \bar q h S$ interaction growing with momentum -- whereas the pure $gg$-induced process exhibits a slight decrease.

If the di-photon resonance is wide, the analyses performed for the $hS$ final state will differ in the details due to changes in the kinematics. We show in Fig.~\ref{fig:ratiowidth} the
variation of the ratio $\rho$ with the width over the mass, $\Gamma / M$, at a fixed mass $M = 750$~GeV, for
$\lambda_{HS} =1$, and for the $b\bar{b}$-initiated process for
$\lambda_{HS} = 0$ (no portal) and $\lambda_{HS} = 1$. One can observe that the central value of
the ratio remains approximately constant in all cases, with only a
slight decrease with increasing width.

In both Figs.~\ref{fig:ratiowidthmass} and~\ref{fig:ratiowidth}, we also provide, as coloured bands, the parton density function uncertainty for the \texttt{MMHT14nlo68cl} set~\cite{Motylinski:2014sya} combined in quadrature with the scale variation between 0.5 and 2.0 times the default central dynamical scale implemented in
\texttt{MadGraph5\_aMC@NLO}.  For a mass of 750 GeV, the total theoretical uncertainties due to scale and PDF
variations are $\sim ^{+40}_{-30} \%$ for the $gg$-induced process, $\sim \pm 10\% $ for the $b\bar{b}$-induced and $\sim \pm 30\%$ for the $s\bar{s}$-induced cases (the latter not
shown in the figure for simplicity).

Assuming a total cross section for the production of a $\gamma\gamma$ resonance of mass $M = 750$~GeV of, say, $\sigma(pp\rightarrow S \rightarrow
\gamma\gamma) = 5$~fb (see below), one would expect a total of $\mathcal{O}(20)$
$hS \rightarrow h\gamma\gamma$ events at the high-luminosity LHC
(HL-LHC, assuming 3000~fb$^{-1}$ of integrated luminosity) if the
process is gluon-fusion initiated and $\mathcal{O}(200)$ events for
$b\bar{b}$-initiated production, for a portal coupling 
$\lambda_{HS} = 1$. Moreover, the minimum expected number of events
for the $b\bar{b}$-initiated process is $\mathcal{O}(80)$, arising for
$\lambda_{HS} \simeq -1.8$ and for the $s\bar{s}$-initiated process
one expects a minimum of $\mathcal{O}(100)$ events for $\lambda_{HS}
\simeq -2.0$. We note here that the positions of the minima for the
$q\bar{q}$-initiated process will change after
cuts due to the varying effect of the analysis on the different pieces
contributing to the cross section.

\begin{figure}[!h]
 \centering
  \includegraphics[width=0.42\textwidth]{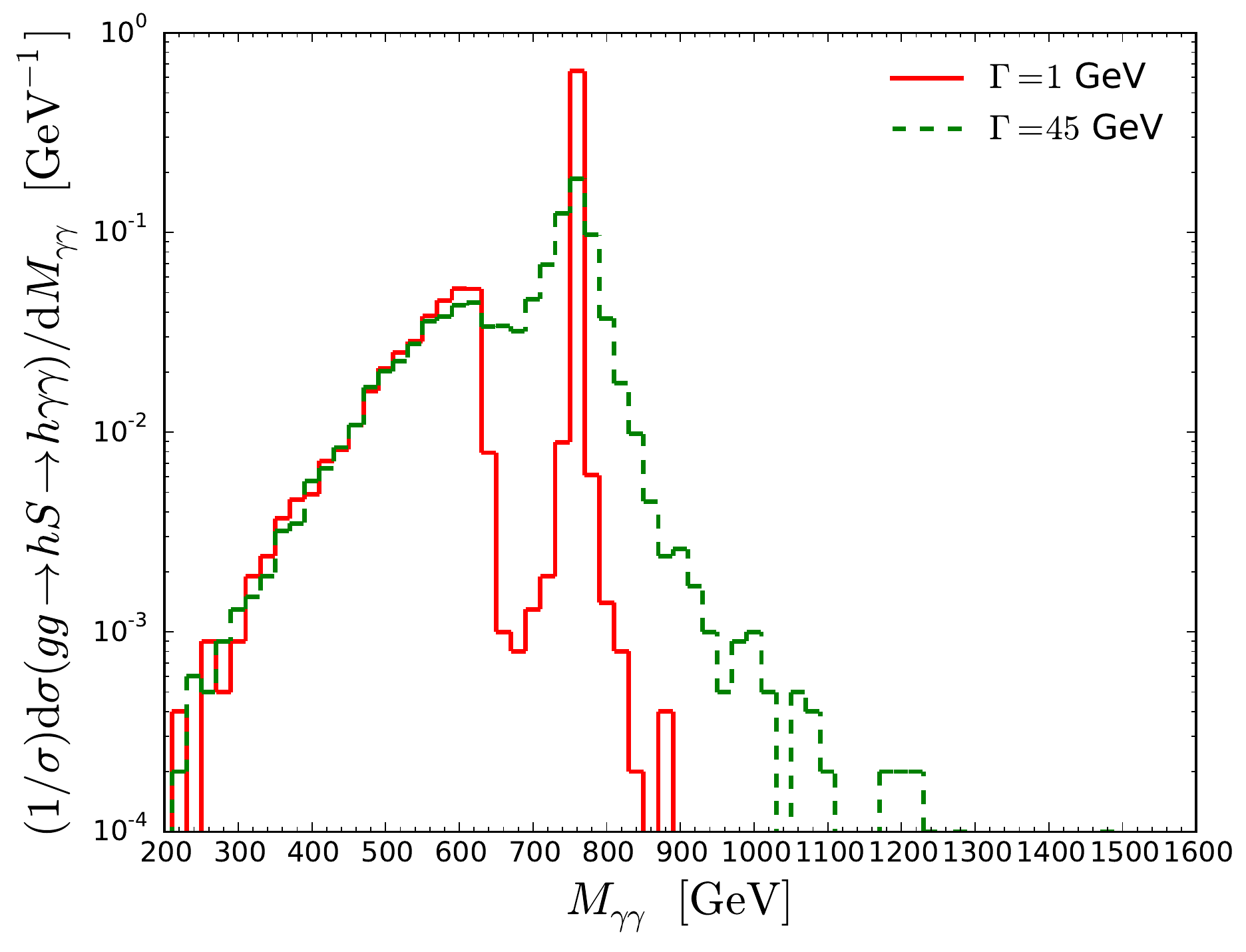}
  \caption{\label{fig:maa_gg} The matrix-element level
    distribution of the di-photon invariant mass, $M_{\gamma\gamma}$,
    in the $gg \rightarrow hS \rightarrow h \gamma\gamma$ process,
    normalised to unity, for the two different width scenarios, $\Gamma =
  1$~GeV and $\Gamma=45$~GeV, for $M=750$~GeV.}
\end{figure}

\begin{figure}[!h]
 \centering
  \includegraphics[width=0.42\textwidth]{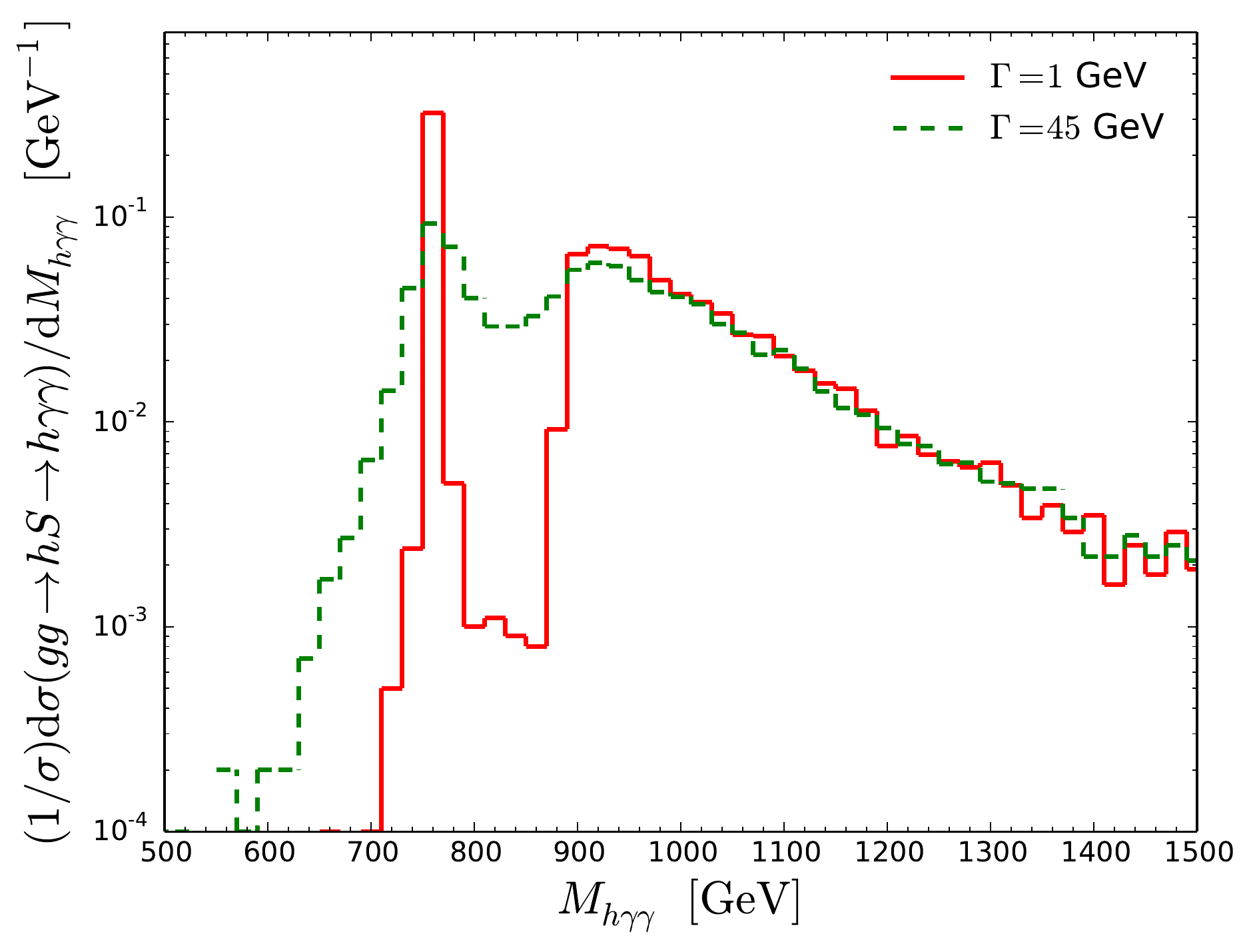}
  \caption{\label{fig:mhaa_gg} The matrix-element-level
    distribution of the combined Higgs boson and di-photon invariant mass, $M_{h\gamma\gamma}$,
   in the $gg \rightarrow hS \rightarrow h \gamma\gamma$ process,
    normalised to unity, for the two different width scenarios, $\Gamma =
  1$~GeV and $\Gamma=45$~GeV, for $M=750$~GeV.}
\end{figure}

The kinematic structure of the $pp \rightarrow hS \rightarrow h \gamma\gamma$ process can be well-described by
examining the distribution of the invariant mass of the $\gamma\gamma$ state, $M_{\gamma\gamma}$, or the distribution of the invariant mass of the Higgs boson and di-photon combination, $M_{h\gamma\gamma}$. In Figs.~\ref{fig:maa_gg} and~\ref{fig:mhaa_gg} we show, respectively, these distributions for the
gluon-fusion-initiated process, for two widths, $\Gamma = 1$~GeV,
$\Gamma = 45$~GeV. For the sake of clarity, here we only show
distributions  for a scalar mass of $M=750$~GeV, but the main features remain
unaltered as long as the scalar is heavier than the Higgs boson, $M>M_h$. The distributions clearly show the existence of two regions: a region in which the intermediate
$s$-channel propagator for the $S$ scalar is on-shell and the final-state $S(\gamma\gamma)$ is off-shell, and a 
region in which the $s$-channel internal propagator is instead off-shell and
the final-state $S(\gamma\gamma)$ is on-shell. The existence of the
former region, $M_{\gamma\gamma} \lesssim M - M_h$, $M_{h\gamma\gamma} \sim M$ which henceforth we will call ``three-body decay'' since the
intermediate $S$ is decaying approximately on-shell, is made possible by the fact that the
mass of the particle produced in association with $S$, the Higgs
boson, is smaller than the masses of $S$ we are considering, $M > M_h=125$~GeV. The other region, $M_{\gamma\gamma} \sim M$, $M_{h\gamma\gamma} \gtrsim M+M_h$ which we will refer to as ``on-shell
di-photon'', exists irrespective of the mass of $S$. Note that both
the three-body decay and on-shell di-photon regions exist even for
$\Gamma/M \ll 1$. The normalised distributions look
identical for \textit{all} values of the portal coupling,
$\lambda_{HS}$ ($\neq 0$), since the dominant contribution stems by far from the
diagram shown in Fig.~\ref{fig:ggdiags}~(a).

\begin{figure}[!h]
 \centering
  \includegraphics[width=0.42\textwidth]{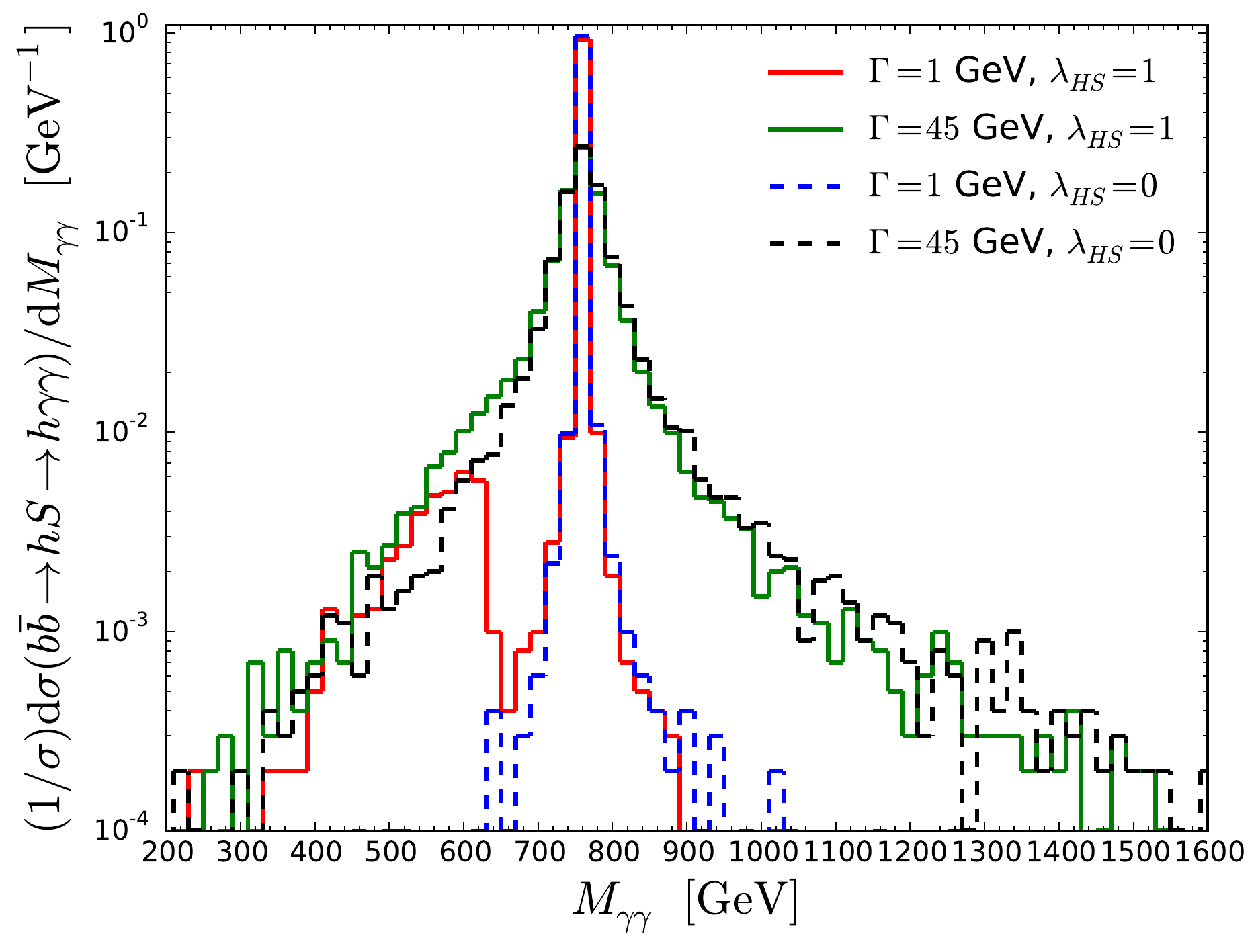}
  \caption{\label{fig:maa_bb} The matrix-element-level
    distribution of the di-photon invariant mass, $M_{\gamma\gamma}$,
    in the $b\bar{b} \rightarrow hS \rightarrow h \gamma\gamma$ process,
    normalised to unity, for two width scenarios, $\Gamma =
  1,~45$~GeV and two values of the portal coupling,
  $\lambda_{HS} = 0,~1$, for $M=750$~GeV. }
\end{figure}

\begin{figure}[!h]
 \centering
  \includegraphics[width=0.42\textwidth]{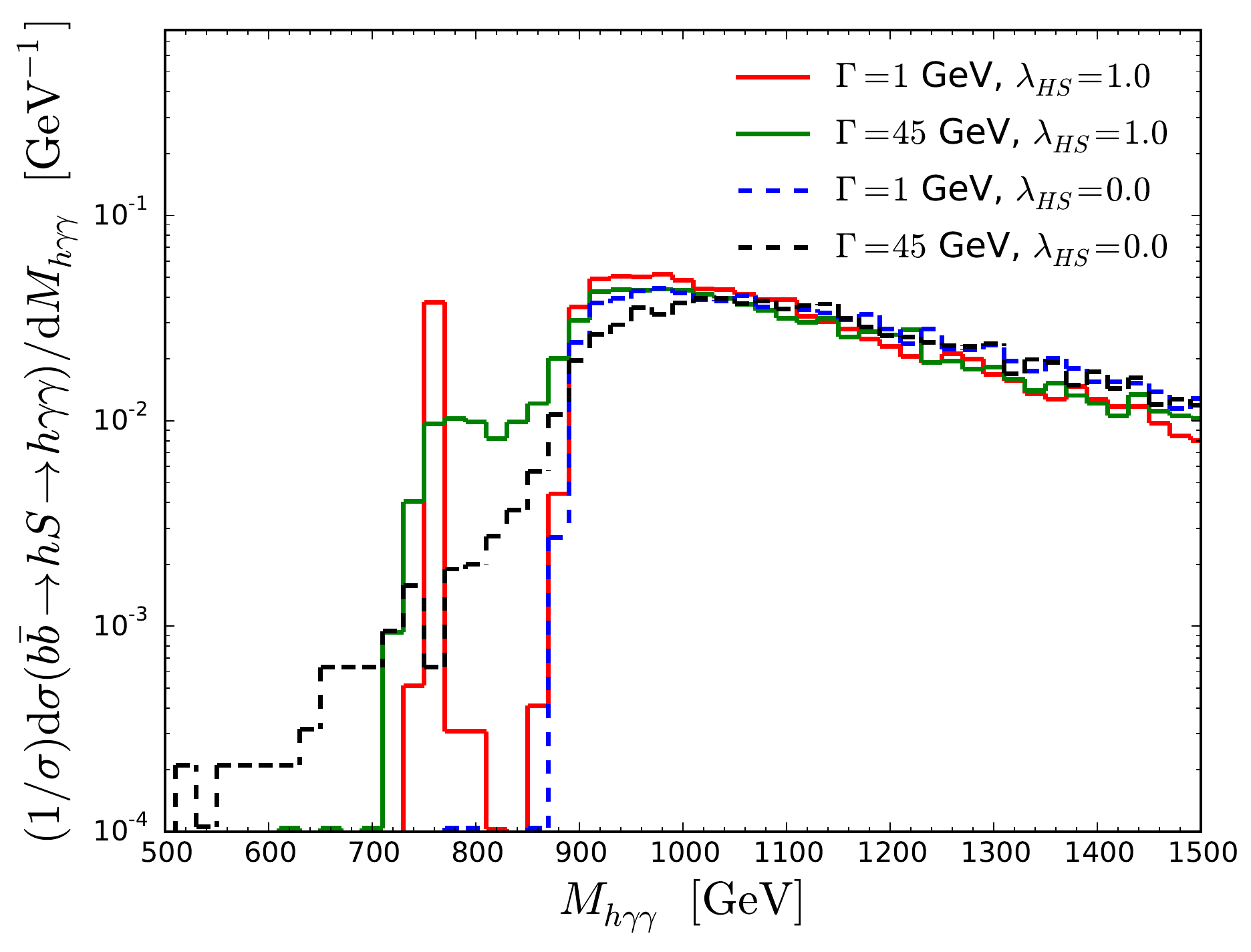}
  \caption{\label{fig:mhaa_bb} The matrix-element level
    distribution of the combined Higgs boson and di-photon invariant mass, $M_{h\gamma\gamma}$,
    in the $b\bar{b} \rightarrow hS \rightarrow h \gamma\gamma$ process,
    normalised to unity, for two width scenarios, $\Gamma =
  1,~45$~GeV and two values of the portal coupling,
  $\lambda_{HS} = 0,~1$, for $M=750$~GeV. }
\end{figure}

In Figs.~\ref{fig:maa_bb} and~\ref{fig:mhaa_bb} we show the di-photon invariant mass and the combined Higgs boson and di-photon invariant mass for the
$b\bar{b}$-initiated process, respectively, for $M=750$~GeV. Evidently the two regions observed for
the $gg$ case are clearly still present for $\lambda_{HS} \neq 0$ and
$\Gamma = 1$~GeV, with the on-shell di-photon region
dominating. 
For $\lambda_{HS} = 0$, the ``three-body decay'' region disappears completely since the resonant
$s$-channel diagram of Fig.~\ref{fig:qqdiags}~(a) vanishes. For large
width the two regions merge into one and the effect of the vanishing
three-body decay region for $\lambda_{HS} =
0$ is not as evident as in the case of small width. The distributions
for the $s\bar{s}$-initiated process exhibit similar features, with different ``mixtures'' between the two regions arising from the differences between the strange and bottom quark parton density functions. We omit them for the sake of simplicity. 

\begin{figure}[!h]
 \centering
  \includegraphics[width=0.42\textwidth]{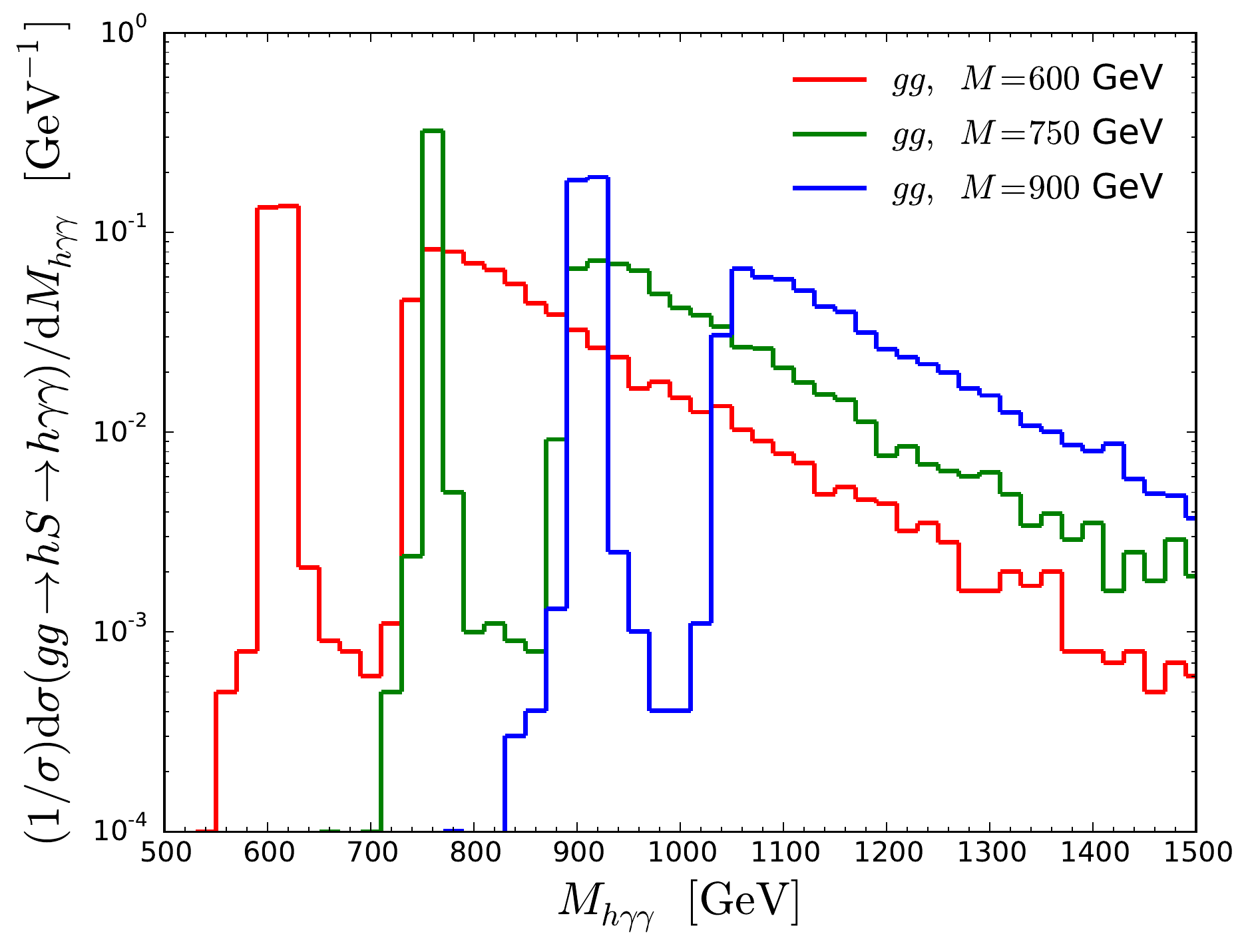}
  \caption{\label{fig:massvar_gg} The matrix-element level
    distribution of the combined Higgs boson and di-photon invariant mass, $M_{h\gamma\gamma}$,
    in the $gg \rightarrow hS \rightarrow h \gamma\gamma$ process,
    normalised to unity, for three benchmark mass scenarios, $M = 600,~750,~900$~GeV, and $\Gamma = 1$~GeV.}
\end{figure}

\begin{figure}[!h]
 \centering
  \includegraphics[width=0.42\textwidth]{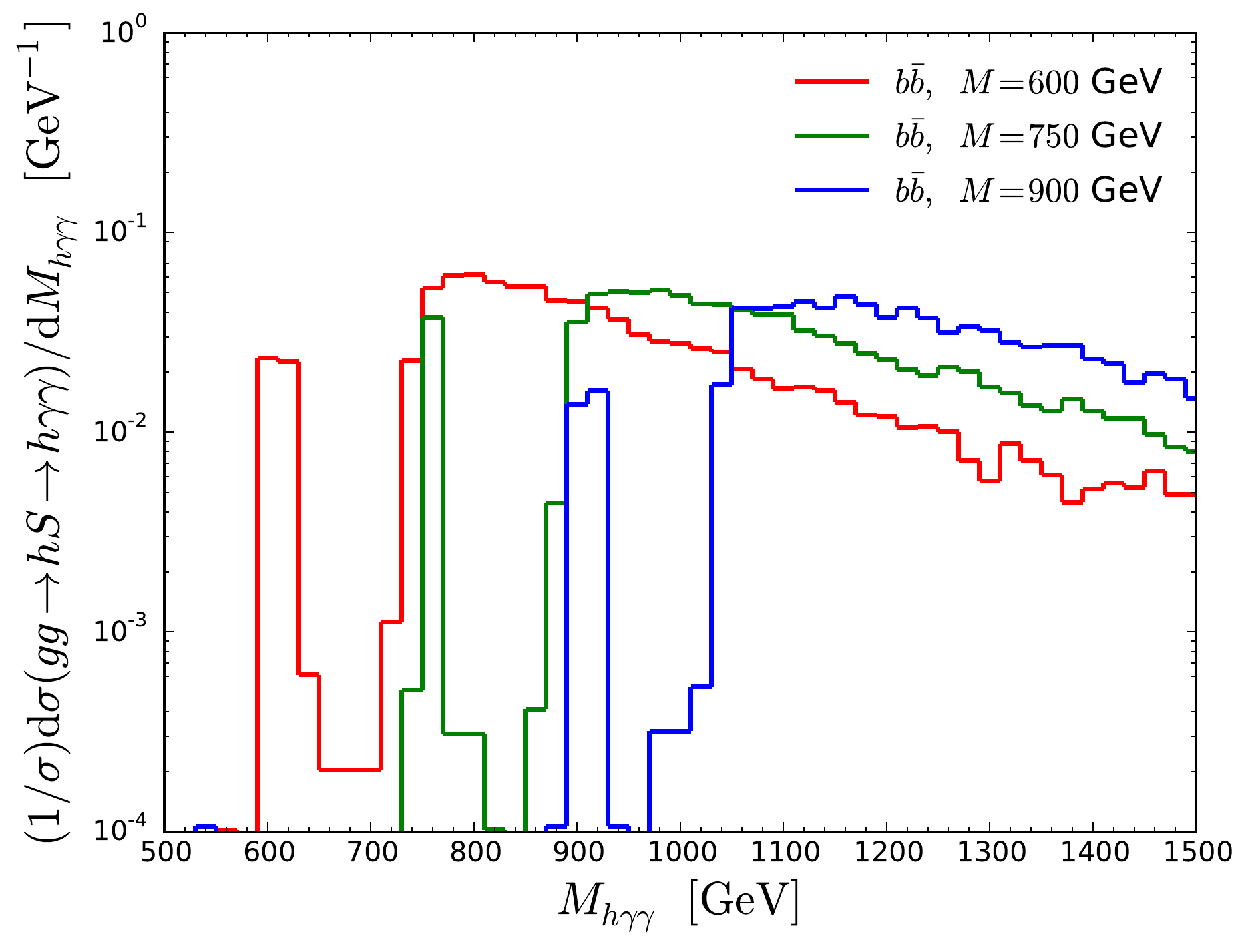}
  \caption{\label{fig:massvar_bb} The matrix-element level
    distribution of the combined Higgs boson and di-photon invariant mass, $M_{h\gamma\gamma}$,
    in the $b\bar{b} \rightarrow hS \rightarrow h \gamma\gamma$ process,
    normalised to unity, for three benchmark mass scenarios, $M =
    600,~750,~900$~GeV, and $\Gamma =1$~GeV, for $\lambda_{HS} = 1$. }
\end{figure}

Figures~\ref{fig:massvar_gg} and~\ref{fig:massvar_bb} show the distributions of the combined invariant mass of the Higgs boson and di-photon, $M_{h\gamma\gamma}$, for the pure $gg$-initiated and pure $b\bar{b}$-initiated cases respectively, for the three values of the scalar mass that we will consider as ``benchmark'' scenarios in our analysis, $M=600,~750,~900$~GeV (see below) and $\Gamma = 1$~GeV. For the $b\bar{b}$ case we only show the $\lambda_{HS} = 1$ distributions for simplicity. They all clearly demonstrate the existence of the main features described for the $M=750$~GeV case, i.e. the ``three-body decay'' and ``on-shell di-photon'' regions.

\section{Event generation and detector simulation.}
\label{sec:sim}
\subsection{Event generation.}

The signal model was generated via an implementation of the
Lagrangian of Eq.~\ref{eq:lagrangian} in
\texttt{FeynRules}~\cite{Christensen:2008py, Alloul:2013bka}. Via the UFO interface~\cite{Degrande:2011ua} this
was used to generate parton-level events employing
\texttt{MadGraph5\_aMC@NLO}~\cite{Alwall:2011uj, Alwall:2014hca}. The
background processes were also generated using
\texttt{MadGraph5\_aMC@NLO}, with appropriate generation-level cuts to
reduce the initial cross sections to a manageable level. All the events were passed through the
\texttt{HERWIG 7}~\cite{Bahr:2008pv, Gieseke:2011na, Arnold:2012fq,
  Bellm:2013lba, Bellm:2015jjp} Monte
Carlo for simulation of the parton shower, the underlying event and
hadronization. As before, the \texttt{MMHT14nlo68cl} PDF set was
employed. To remain conservative, we consider collisions at the LHC at a centre-of-mass energy of 13~TeV. The possible increase of energy to 14~TeV will increase rates in the considered processes by $\mathcal{O}(10\%)$.

Since we expect to impose cuts on the di-photon mass window,
$M_{\gamma\gamma}$, that are sufficiently far away from the Higgs boson
resonance, we can immediately exclude any background processes
containing $h\rightarrow \gamma\gamma$ from the analysis. For this
reason we do not include associated Higgs boson production with a vector
boson or Higgs boson pair
production, $t\bar{t}h$ production, and so on. This implies that the
relevant backgrounds are those with non-resonant $\gamma\gamma$
production, other processes that involve $S\rightarrow \gamma\gamma$, and reducible backgrounds. We thus consider the following
processes: $\gamma+$jets, $\gamma\gamma+$jets, events with at least one true
$b$-quark at parton-level ($b$+jets), $Z\gamma\gamma$ with
$Z\rightarrow b\bar{b}$, $t\bar{t}\gamma\gamma$ including all the
decay modes of the top quarks and the production of the resonance $S$ in association with a non-resonant $b\bar{b}$ pair.\footnote{We also considered the $h\gamma\gamma$ process, including the loop-induced pieces~\cite{Hirschi:2015iia}, but found that it possess a negligible cross section.} All the multi-jet
processes are generated without merging to the parton shower, in the
five-flavour scheme, with four outgoing partons at the matrix-element level. 

The calculation of higher-order QCD corrections to
these multi-leg processes, particularly when restricting the phase space with cuts,
is numerically challenging at present. To remain conservative, we will assume that
the corrections are large and apply $K$-factors of $K=2$ to all the
background processes. For the signal and the $b\bar{b}S$ associated production we do not apply any $K$-factors
since the corrections are approximately absorbed into the ratio with the single inclusive production of the $S$ resonance, see below. Throughout this article we assume that  $\sigma(pp \rightarrow S \rightarrow \gamma\gamma) = 10,~5,~1$~fb, corresponding to the benchmark masses $M=600,~750,~900$~GeV, fixing the product $c_G^S c_{\gamma}^S$ (or 
$(y_d^S)^{ii} c_{\gamma}^S$), which drops out in the ratio~$\rho$. The values of the cross sections are motivated by the current ATLAS~\cite{ATLAS:2016eeo} and CMS~\cite{Khachatryan:2016yec} limits on di-photon resonances.

Note that it turns out that the non-resonant $b\bar{b}S$ process is only relevant for gluon-fusion production of $S$, and we only report numbers for that in what follows. 

\subsection{Detector simulation.}

In the hadron-level analysis that follows, performed without using any dedicated detector simulation software, we consider all particles within a pseudo-rapidity of
$|\eta| < 5$ and $p_T > 100$~MeV. We smear the momenta of all
reconstructed objects (i.e. jets, electrons, muons and photons) according
to HL-LHC projections~\cite{TheATLAScollaboration:performance1, TheATLAScollaboration:performance2}. We also apply the
relevant reconstruction efficiencies. We simulate $b$-jet tagging by
looking for jets containing $B$-hadrons, that we have set to stable in
the simulation, and considering them as the
$b$-jet candidates. The mis-tagging of $c$-jets to $b$-jets is performed by choosing
$c$-jet candidates (after hadronization) as those jets that lie within a distance $\Delta R < 0.4$ from
$c$-quarks (after the parton shower), with transverse momentum $p_T > 1$~GeV.\footnote{This procedure of associating jets to $c$-quarks is expected to be conservative.} We apply a flat b-tagging efficiency of 70\% and a
mis-tag rate of 1\% for light-flavour jets and 10\% for
charm-quark-initiated jets.

We reconstruct jets using the anti-$k_t$ algorithm available in the
\texttt{FastJet} package~\cite{Cacciari:2011ma, Cacciari:2005hq}, with a radius
parameter of $R=0.4$. We only consider jets, photons and leptons with $p_T > 30$~GeV within
pseudo-rapidity $|\eta| < 2.5$ in our analysis. The jet-to-lepton mis-identification probability is
taken to be $\mathcal{P}_{j\rightarrow \ell} = 0.0048 \times
\mathrm{e}^{-0.035 p_{Tj}/\mathrm{GeV}}$ and the jet-to-photon
mis-identification probability was taken to be $\mathcal{P}_{j\rightarrow \gamma} = 0.0093 \times
 \mathrm{e}^{-0.036 p_{Tj}/\mathrm{GeV}}$~\cite{TheATLAScollaboration:performance1, TheATLAScollaboration:performance2}, both flat in pseudo-rapidity. We demand all leptons and
 photons to be isolated, where an isolated object is defined to have $\sum_i p_{T,i}$
less than 15\% of its transverse momentum in a cone of $\Delta R = 0.2$ around it. 

\section{Detailed analysis.}
\label{sec:analysis}

We consider events with two reconstructed $b$-jets and two isolated photons as
defined in Section~\ref{sec:sim}. Note that this final state has been previously considered in the context of searches for Higgs boson pair production, e.g. in~\cite{Baur:2003gp, Baglio:2012np, Yao:2013ika, Barger:2013jfa, Chen:2013emb}. We impose the following `acceptance' cuts to all samples:

\begin{itemize}
\item $b$-jets: transverse momenta $p_{T,b1} > 30$~GeV, $p_{T,b2} >
  30$~GeV, all $b$-jets within $|\eta| < 2.5$,
\item photons: transverse momenta $p_{T,\gamma 1} > 30$~GeV,
  $p_{T,\gamma 2} > 30$~GeV, all photons within $|\eta| < 2.5$,
\item invariant mass of the two $b$-jets $M_{b\bar{b}} \in [90,
    160]$~GeV,
\item invariant mass of the two photons $M_{\gamma\gamma} > M-300$~GeV,
\item veto events with leptons of $p_T > 25$~GeV within $|\eta| <
  2.5$, 
\end{itemize}
for each of the considered di-photon resonance masses, $M$. 

\begin{table}[!t]
\begin{ruledtabular}
\begin{tabular}{cc}
process & acceptance $\sigma$~[fb] \\\hline
$gg \rightarrow h(b\bar{b}) S(\gamma\gamma)$, $\Gamma = 1$~GeV, $\lambda_{HS} = 1$ & 0.00054 \\ 
$gg \rightarrow h(b\bar{b}) S(\gamma\gamma)$, $\Gamma = 45$~GeV, $\lambda_{HS} = 1$ & 0.00055 \\\hline
$b\bar{b} \rightarrow h(b\bar{b}) S(\gamma\gamma)$, $\Gamma = 1$~GeV, $\lambda_{HS} = 1$ & 0.00266  \\
$b\bar{b} \rightarrow h(b\bar{b}) S(\gamma\gamma)$, $\Gamma = 45$~GeV, $\lambda_{HS} = 1$ & 0.00254 \\\hline
$b\bar{b} \rightarrow h(b\bar{b}) S(\gamma\gamma)$, $\Gamma = 1$~GeV, $\lambda_{HS} = 0$ & 0.00184 \\
$b\bar{b}  \rightarrow h(b\bar{b}) S(\gamma\gamma)$, $\Gamma = 45$~GeV, $\lambda_{HS} = 0$ &  0.00172\\\hline
$s\bar{s} \rightarrow h(b\bar{b}) S(\gamma\gamma)$, $\Gamma = 1$~GeV, $\lambda_{HS} = 1$ & 0.00366  \\
$s\bar{s} \rightarrow h(b\bar{b}) S(\gamma\gamma)$, $\Gamma = 45$~GeV, $\lambda_{HS} = 1$ &  0.00370 \\\hline
$s\bar{s} \rightarrow h(b\bar{b}) S(\gamma\gamma)$, $\Gamma = 1$~GeV, $\lambda_{HS} = 0$ & 0.00291 \\
$s\bar{s}  \rightarrow h(b\bar{b}) S(\gamma\gamma)$, $\Gamma = 45$~GeV, $\lambda_{HS} = 0$ &  0.00249\\\hline
at least one $b$-quark + jets & 0.31300  \\
$\gamma$ + jets & 0.11259\\
$\gamma\gamma$ + jets & 0.15766 \\
$Z\gamma\gamma \rightarrow (b\bar{b})\gamma\gamma$ & 0.00489 \\
$t\bar{t} \gamma\gamma$ & 0.00281 \\\hline
$gg \rightarrow b\bar{b} S(\gamma\gamma)$, $\Gamma = 1$~GeV & 0.00058  \\ 
$gg \rightarrow b\bar{b} S(\gamma\gamma)$, $\Gamma = 45$~GeV & 0.00063\\ 
\end{tabular}
\caption{
\label{tb:acceptance}
The expected cross sections at 13 TeV $pp$ collision energy for all the considered processes after
  acceptance cuts for $M=750$~GeV and $\Gamma = 1,~45$~GeV. All branching ratios, acceptances and tagging rates
  have been applied. We have assumed that the single production cross
  section for a di-photon scalar resonance of $M=750$~GeV is $\sigma(pp \rightarrow S \rightarrow \gamma\gamma) =
  5$~fb.}
\end{ruledtabular}
\end{table}

The cross sections after application of the acceptance cuts are given in
Table~\ref{tb:acceptance} for two values of the widths $\Gamma =
1$~GeV and $\Gamma = 45$~GeV and for $M=750$~GeV. For the case of $q\bar{q}$ we consider as examples $\lambda_{HS} = 1$ and $\lambda_{HS} = 0$. Throughout this analysis, the total signal cross section was calculated by using the ratio $\rho$ (derived in Section \ref{sec:hS}) as $\sigma(pp \rightarrow hS \rightarrow h\gamma\gamma) = \rho \times \sigma(pp \rightarrow S \rightarrow \gamma\gamma)$, where $\sigma(pp \rightarrow S \rightarrow \gamma\gamma) = 10,~5,~1$~fb for $M = 600,~750,~900$~GeV, and including the decay $h \to b \bar b$. This cross section was employed as the normalisation of the signal event samples (before analysis cuts). The expected number of signal events, for $\lambda_{HS} \sim \mathcal{O}(1)$, after acceptance cuts is $\mathcal{O}(1)-\mathcal{O}(10)$ at 3000~fb$^{-1}$ of integrated luminosity. However, as already discussed, one should keep in mind that the cross
section grows with $\lambda_{HS}^2$ in both $gg$- and
$q\bar{q}$-initiated production.

\begin{figure}[!h]
 \centering
  \includegraphics[width=0.42\textwidth]{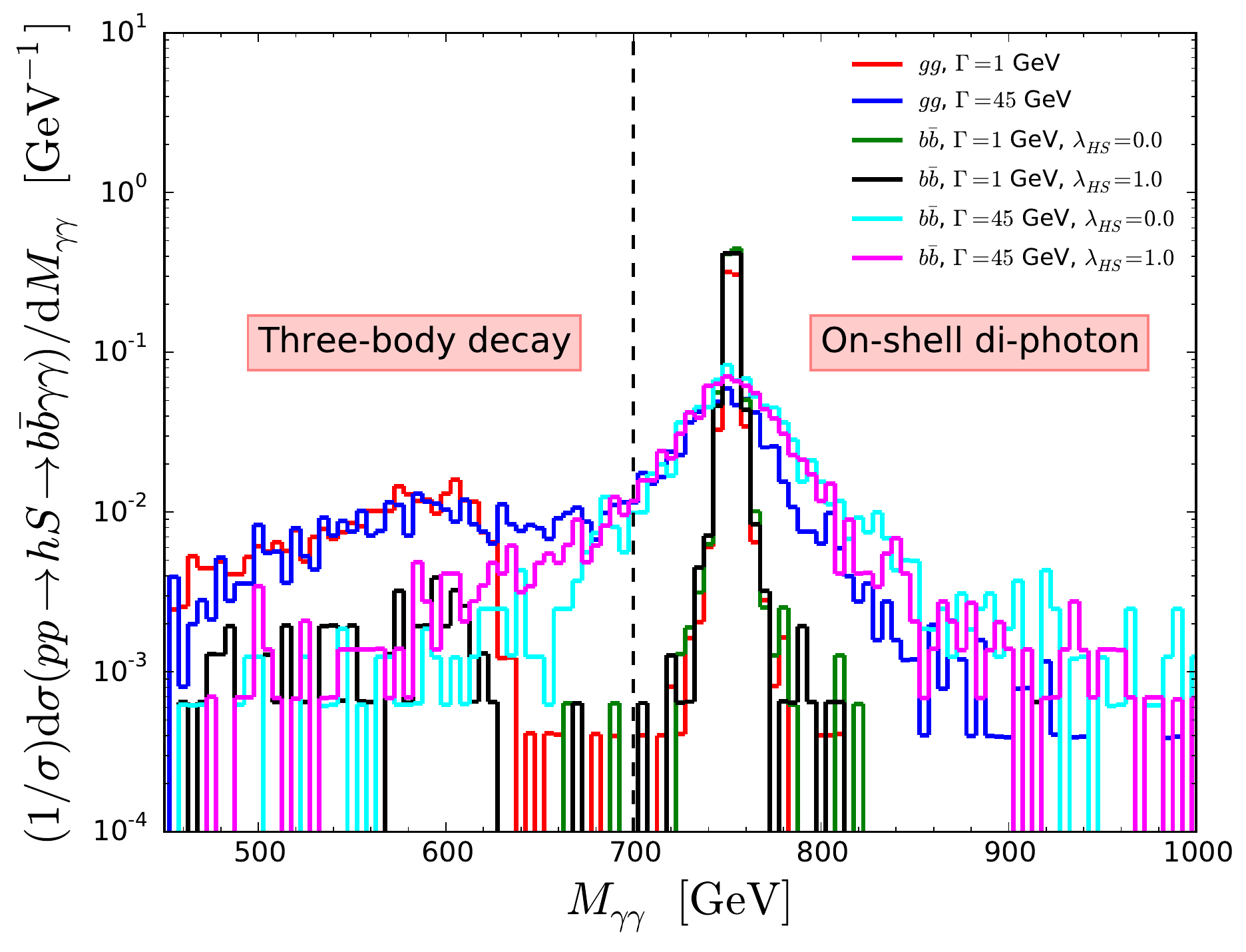}
  \caption{\label{fig:maa_gg_analy} The distribution of the di-photon invariant mass, $M_{\gamma\gamma}$,
    for the $gg \rightarrow hS \rightarrow b \bar b \gamma\gamma$ process
    after acceptance cuts, normalised to unity, for two width scenarios, $\Gamma =
  1$~GeV and $\Gamma=45$~GeV, while $M=750$~GeV. }
\end{figure}

\begin{figure}[!h]
 \centering
  \includegraphics[width=0.42\textwidth]{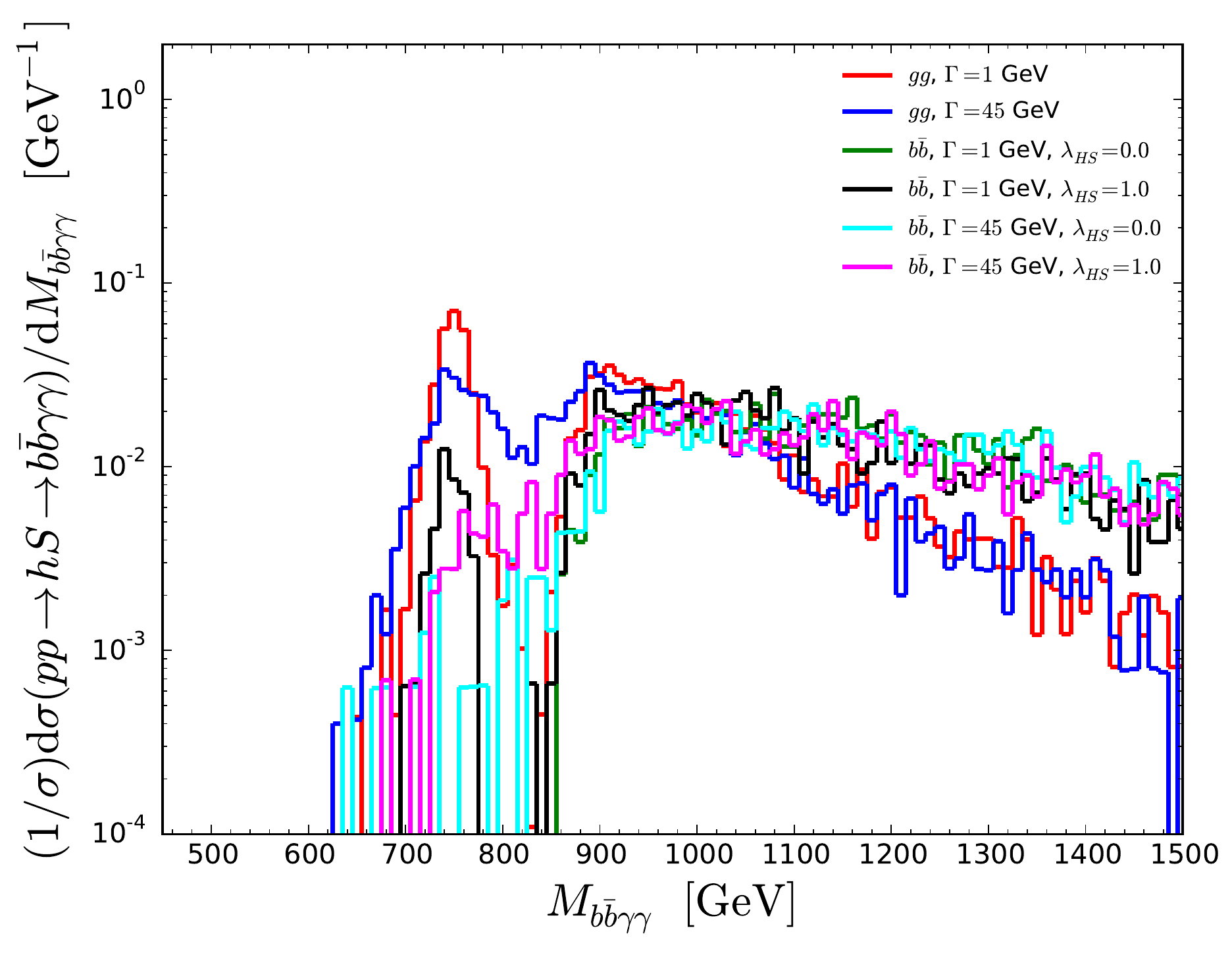}
  \caption{\label{fig:mhaa_gg_analy} The distribution of the combined di-photon + $b \bar b$ invariant mass, $M_{b \bar b \gamma\gamma}$,
    for the $gg \rightarrow hS \rightarrow b \bar b \gamma\gamma$ process
    after acceptance cuts, normalised to unity, for two width scenarios, $\Gamma =
  1$~GeV and $\Gamma=45$~GeV, while $M=750$~GeV. }
\end{figure}

The resulting di-photon invariant mass after acceptance cuts is shown in
Fig.~\ref{fig:maa_gg_analy} for the example of $M=750$~GeV. The $M_{\gamma\gamma}$ observable can be
used to separate the analysis into the two regions described
in Section~\ref{sec:hS}: the ``three-body decay'' region (``TBD'') and the
``on-shell di-photon'' region (``OS$\gamma\gamma$''). The separation is identical in both $gg$- and $q\bar{q}$-initiated processess. We choose: $M_{\gamma\gamma} < M-50$~GeV for the ``TBD'' region and $M_{\gamma\gamma} > M-50$~GeV for the ``OS$\gamma\gamma$'' region for a di-photon resonance mass, $M$. We also show the distribution of the combined invariant mass of the two $b$-jet candidates and the di-photon system, $M_{b\bar{b}\gamma\gamma}$ in Fig.~\ref{fig:mhaa_gg_analy}, which also clearly demonstrates the existence of the two regions. 

We apply further cuts to improve signal and background discrimination. As we did not attempt to fully optimize the cuts in the present analysis, we apply a common set of cuts along with invariant mass cuts on the observables $M_{\gamma\gamma}$ and $M_{b\bar{b}\gamma\gamma}$ that provide the main distinction between the two regions. The common cuts applied in each region are shown in Table~\ref{tb:commoncuts} and the specific invariant mass cuts are shown in Table~\ref{tb:masscuts}. Effectively, the cuts aim to exploit the fact that the photons in the signal are harder than in the backgrounds and also feature tighter di-photon and $b\bar{b}$ mass windows, particularly in the ``OS$\gamma\gamma$'' region for the former. 

\begin{table}[!h]
\begin{ruledtabular}
\begin{tabularx}{\linewidth}{*{2}{p{.49\linewidth}}}
\textbf{observable} & \textbf{cut}\\ \hline
$p_{T,\gamma 1}$ & $> 200$~GeV \\ 
$p_{T,\gamma 2}$ & $> 120$~GeV \\ 
$\Delta R(\gamma,\gamma)$ &$\in [2.0, 4.0]$\\
$M_{b\bar{b}} $ &  $\in [100, 150]$~GeV \\ 
$\Delta R(b,\bar{b})$ & $\in [0.8, 3.0]$ \\
$\Delta R (\gamma\gamma,b_2) $ & $< 3.0$ \\ 
\end{tabularx}
\caption{\label{tb:commoncuts} The additional common cuts applied along with the acceptance cuts in both the ``three-body decay'' 
and the ``on-shell di-photon'' region for the
$gg$- and $q\bar{q}$-initiated processess. The labels
``$1$'' and ``$2$'' correspond to the hardest and
second hardest reconstructed objects (photons or $b$-jets), respectively.}
\end{ruledtabular}
\end{table}

\begin{table}[!h]
\begin{ruledtabular}
\begin{tabular}{ccc}
 ~ & \textbf{``TBD'}  & \textbf{``OS$\gamma\gamma$''}\\\hline
$\Gamma =1$~GeV &  \hfill & \hfill\\
$M_{\gamma\gamma} $ & $\in M_{-300}^{-110}$~GeV & $\in M\pm 5$~GeV \\ 
$M_{b\bar{b}\gamma\gamma} $ & $\in M \pm 30$~GeV & - \\ \hline
$\Gamma =45$~GeV: & &  \\
$M_{\gamma\gamma} $  & $\in M^{-110}_{-300}$~GeV& $\in M \pm 40$~GeV\\ 
$M_{b\bar{b}\gamma\gamma}$ & $\in M \pm 40$~GeV & - \\ 
\end{tabular}
\caption{\label{tb:masscuts} The additional invariant mass cuts applied along with the further cuts of Table~\ref{tb:commoncuts}, in the ``three-body decay'' region (``TBD'') and the ``on-shell di-photon'' region (``OS$\gamma\gamma$''), for both the $gg$- and $q\bar{q}$-initiated processess for a scalar di-photon resonance of mass $M$. The different choices for the mass windows were made according to the width of the resonance, $\Gamma$.}
\end{ruledtabular}
\end{table}

\begin{table}[!t]
\begin{ruledtabular}
\begin{tabular}{ccc}
process  & ``TBD''  & ``OS$\gamma\gamma$''\\\hline
$gg \rightarrow h(b\bar{b}) S(\gamma\gamma)$, $\Gamma = 1$~GeV, $\lambda_{HS} = 1$ & 0.00007  &   0.00020 \\ 
$gg \rightarrow h(b\bar{b}) S(\gamma\gamma)$, $\Gamma = 45$~GeV, $\lambda_{HS} = 1$ & 0.00006 &   0.00014 \\\hline
$b\bar{b} \rightarrow h(b\bar{b}) S(\gamma\gamma)$, $\Gamma = 1$~GeV, $\lambda_{HS} = 1$ & 0.00007 & 0.00105  \\
$b\bar{b} \rightarrow h(b\bar{b}) S(\gamma\gamma)$, $\Gamma = 45$~GeV, $\lambda_{HS} = 1$ &0.00038  & 0.00074 \\\hline
$b\bar{b} \rightarrow h(b\bar{b}) S(\gamma\gamma)$, $\Gamma = 1$~GeV, $\lambda_{HS} = 0$ & 0  &  0.00064\\
$b\bar{b}  \rightarrow h(b\bar{b}) S(\gamma\gamma)$, $\Gamma =  45$~GeV, $\lambda_{HS} = 0$ & 0.00018 & 0.00055\\\hline
$s\bar{s} \rightarrow h(b\bar{b}) S(\gamma\gamma)$, $\Gamma = 1$~GeV, $\lambda_{HS} = 1$ &0.00005  &  0.00134\\
$s\bar{s} \rightarrow h(b\bar{b}) S(\gamma\gamma)$, $\Gamma = 45$~GeV, $\lambda_{HS} = 1$ & 0.00005 & 0.00107 \\\hline
$s\bar{s} \rightarrow h(b\bar{b}) S(\gamma\gamma)$, $\Gamma = 1$~GeV, $\lambda_{HS} = 0$ &  0& 0.00101\\
$s\bar{s}  \rightarrow h(b\bar{b}) S(\gamma\gamma)$, $\Gamma = 45$~GeV, $\lambda_{HS} = 0$ & 0.00003&  0.00072\\\hline
at least one $b$-quark + jets &  $\mathcal{O}(10^{-7})$  & $\mathcal{O}(10^{-10})$ \\
$\gamma$ + jets & 0.00025 & $\mathcal{O}(10^{-7})$  \\
$\gamma\gamma$ + jets & 0.00182  & 0.00003 \\
$Z\gamma\gamma \rightarrow (b\bar{b})\gamma\gamma$ & $6\times 10^{-5}$ & $10^{-5}$ \\
$t\bar{t} \gamma\gamma$ & $9\times10^{-5}$ & $ 10^{-5}$ \\ \hline
$gg \rightarrow b\bar{b} S(\gamma\gamma)$, $\Gamma = 1$~GeV& $\lesssim \mathcal{O}(10^{-6})$ & 0.00020 \\ 
$gg \rightarrow b\bar{b} S(\gamma\gamma)$, $\Gamma = 45$~GeV& $\mathcal{O}(10^{-6})$  & 0.00018\\ 
\end{tabular}
\caption{\label{tb:furthercuts} The expected cross sections at 13 TeV $pp$ collision energy in fb for all the considered processes after the application of further cuts as in Tables~\ref{tb:commoncuts} and~\ref{tb:masscuts}, for $M=750$~GeV and $\Gamma = 1,~45$~GeV. All branching ratios, acceptances and tagging rates have been applied. We have assumed that the single production cross
  section for a di-photon scalar resonance of $M=750$~GeV is $\sigma(pp \rightarrow S \rightarrow \gamma\gamma) = 5$~fb.}
\end{ruledtabular}
\end{table}

We show the resulting cross sections after the application of these further cuts in Table~\ref{tb:furthercuts}, for the case of $M=750$~GeV. A high efficiency is maintained for the signal, with high rejection factors for the background processes. We note again that the $b\bar{b}S$ associated production process is relevant only for the gluon-fusion scenario. 

To obtain the 95\% confidence-level exclusion regions for
$\lambda_{HS}$ we use Poissonian statistics to calculate the
probabilities. Since we have assumed that the production of a scalar
di-photon resonance will have been observed, we have to construct a null
hypothesis compatible with such an observation providing the {\it expected} number of events at the LHC, that we will
confront with the theory predictions in the parameter space to be tested.
If these numbers differ by a certain significance, the corresponding point is expected to be excluded with this significance.
In particular, any hypothesis has to be realistic and remain within the bounds of our model. Our underlying assumption is thus chosen to be that the
scalar resonance $S$ is produced purely in gluon fusion to good approximation and that there is no portal coupling, $\lambda_{HS}=0$, which means there is basically no $h+S$ associated production.
For further technical details on this statistical procedure, see Appendix C of
Ref.~\cite{Papaefstathiou:2014oja}. We do not incorporate the effect of
systematic uncertainties on the signal or backgrounds. To perform a combination of the two analysis regions, ``TBD'' and
``$OS\gamma\gamma$'', we employ the ``Stouffer method''~\cite{stouffer}, where the combined significance, $\Omega$, is given, in terms of the individual significances $\Omega_i$, as:
\begin{equation}
\Omega = \frac{1}{\sqrt{k}} \sum_{i=1}^k \Omega_i\;.
\end{equation}

\begin{figure}[!h]
 \centering
  \includegraphics[width=0.42\textwidth]{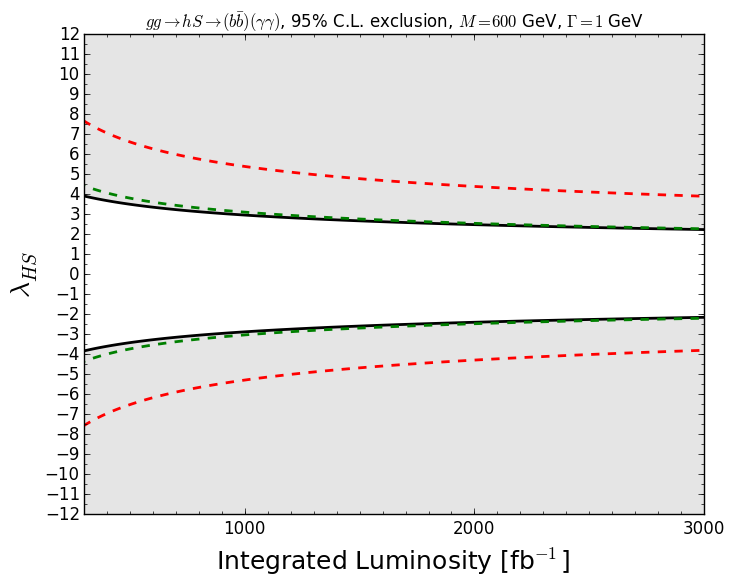}
  \caption{\label{fig:exc_gg_600_1} The grey-shaded area shows the 95\% confidence-level exclusion region for the portal coupling $\lambda_{HS}$ for the $gg$-induced case and $M=600$~GeV, $\Gamma = 1$~GeV, for the combination of the ``TBD'' (red) and ``OS$\gamma\gamma$'' (green) regions as defined in the main text. We have assumed that the single production cross section $\sigma(pp \rightarrow S \rightarrow \gamma\gamma) = 10$~fb.}
\end{figure}

\begin{figure}[!h]
 \centering
  \includegraphics[width=0.42\textwidth]{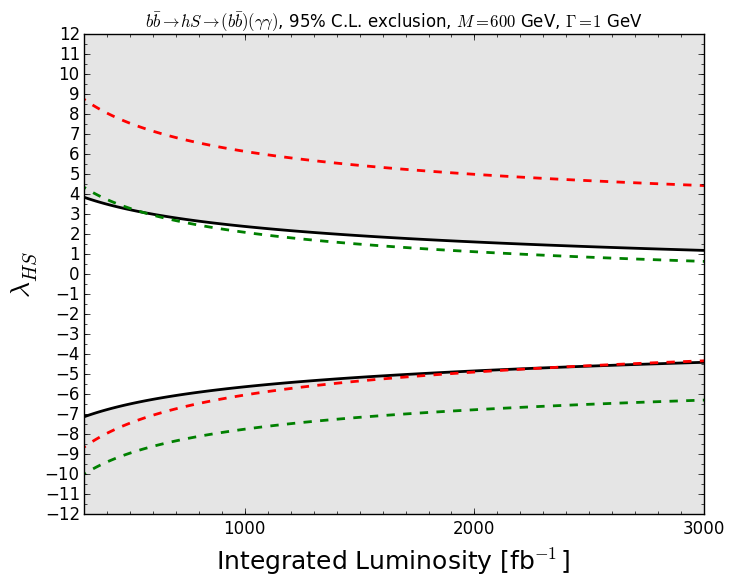}
  \caption{\label{fig:exc_bb_600_1} The grey-shaded area shows the 95\% confidence-level exclusion region for the portal coupling $\lambda_{HS}$ for the $b\bar{b}$-induced case and $M=600$~GeV, $\Gamma = 1$~GeV, for the combination of the ``TBD'' (red) and ``OS$\gamma\gamma$'' (green) regions as defined in the main text. We have assumed that the single production cross section $\sigma(pp \rightarrow S \rightarrow \gamma\gamma) = 10$~fb.}
\end{figure}

\begin{figure}[!h]
 \centering
  \includegraphics[width=0.42\textwidth]{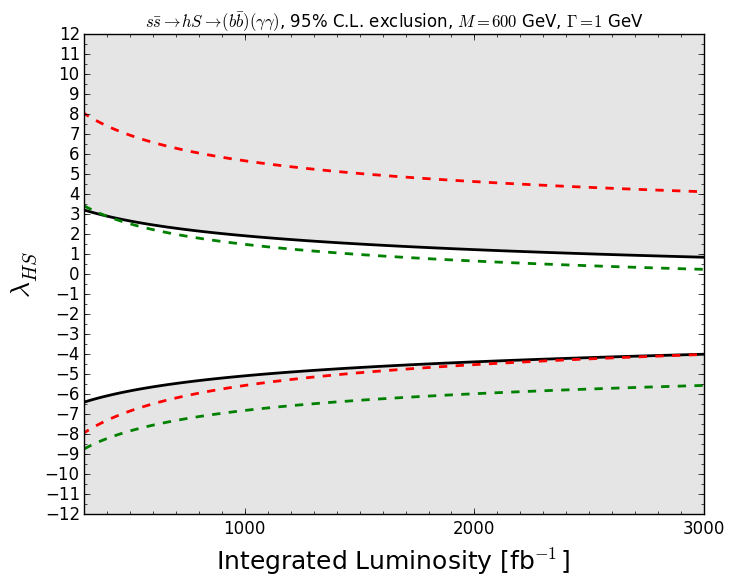}
  \caption{\label{fig:exc_ss_600_1} The grey-shaded area shows the 95\% confidence-level exclusion region for the portal coupling $\lambda_{HS}$ for the $s\bar{s}$-induced case and $M=600$~GeV, $\Gamma = 1$~GeV, for the combination of the ``TBD'' (red) and ``OS$\gamma\gamma$'' (green) regions as defined in the main text. We have assumed that the single production cross section $\sigma(pp \rightarrow S \rightarrow \gamma\gamma) = 10$~fb.}
\end{figure}

\begin{figure}[!h]
 \centering
  \includegraphics[width=0.42\textwidth]{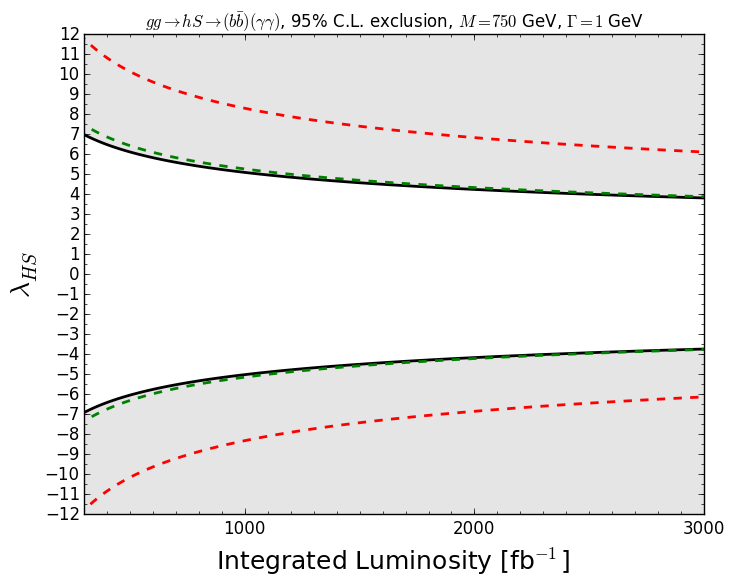}
  \caption{\label{fig:exc_gg_1} The grey-shaded area shows the 95\% confidence-level exclusion region for the portal coupling $\lambda_{HS}$ for the $gg$-induced case and $M=750$~GeV, $\Gamma = 1$~GeV, for the combination of the ``TBD'' (red) and ``OS$\gamma\gamma$'' (green) regions as defined in the main text. We have assumed that the single production cross section $\sigma(pp \rightarrow S \rightarrow \gamma\gamma) = 5$~fb.}
\end{figure}

\begin{figure}[!h]
 \centering
  \includegraphics[width=0.42\textwidth]{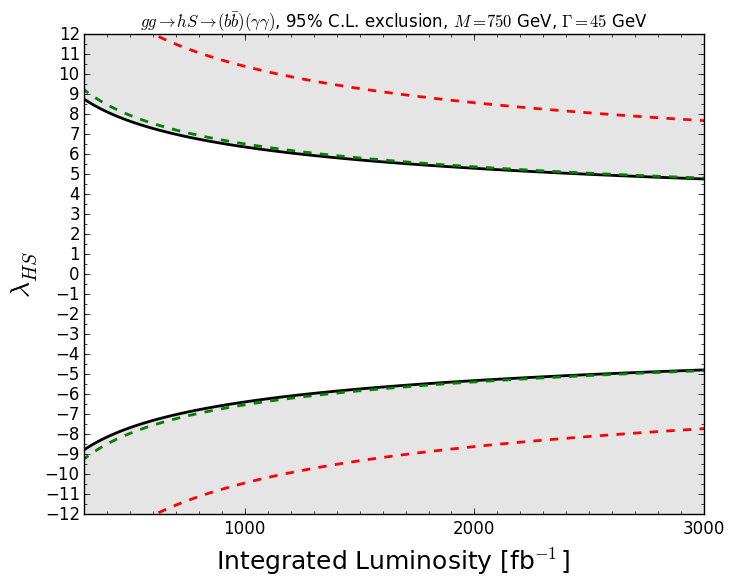}
  \caption{\label{fig:exc_gg_45} The grey-shaded area shows the 95\% confidence-level exclusion region for the portal coupling $\lambda_{HS}$ for the $gg$-induced case and $M=750$~GeV, $\Gamma = 45$~GeV, for the combination of the ``TBD'' (red) and ``OS$\gamma\gamma$'' (green) regions as defined in the main text. We have assumed that the single production cross section $\sigma(pp \rightarrow S \rightarrow \gamma\gamma) = 5$~fb.}
\end{figure}

\begin{figure}[!h]
 \centering
  \includegraphics[width=0.42\textwidth]{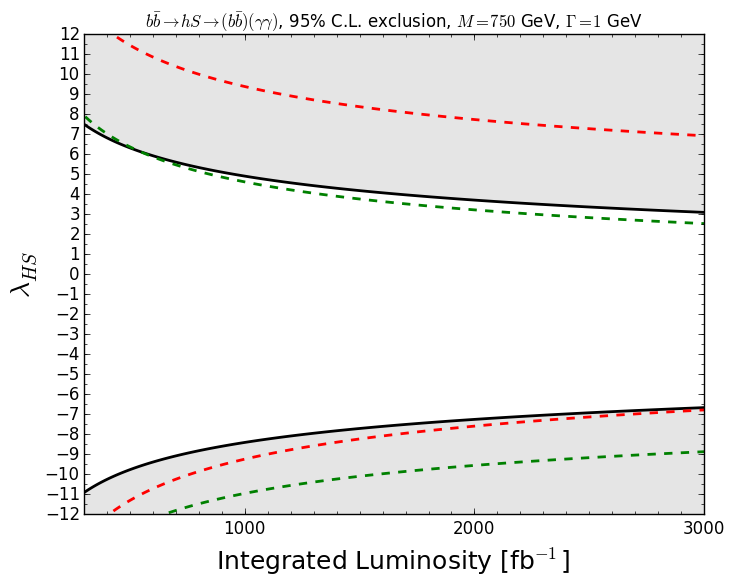}
  \caption{\label{fig:exc_bb_1}  The grey-shaded area shows the 95\% confidence-level exclusion region for the portal coupling $\lambda_{HS}$ for the $b\bar{b}$-induced case and $M=750$~GeV, $\Gamma = 1$~GeV, for the combination of the ``TBD'' (red) and ``OS$\gamma\gamma$'' (green) regions as defined in the main text. We have assumed that the single production cross section $\sigma(pp \rightarrow S \rightarrow \gamma\gamma) = 5$~fb.}
\end{figure}

\begin{figure}[!h]
 \centering
  \includegraphics[width=0.42\textwidth]{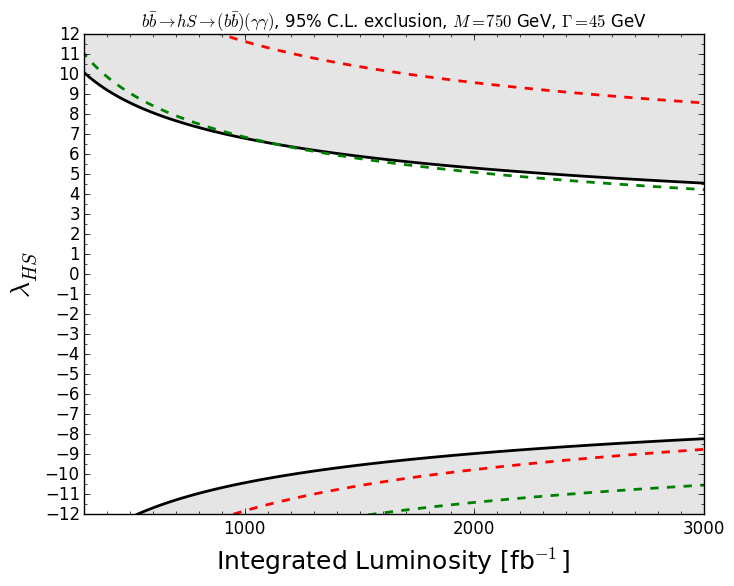}
  \caption{\label{fig:exc_bb_45}  The grey-shaded area shows the 95\% confidence-level exclusion region for the portal coupling $\lambda_{HS}$ for the $b\bar{b}$-induced case and $M=750$~GeV, $\Gamma = 45$~GeV, for the combination of the ``TBD'' (red) and ``OS$\gamma\gamma$'' (green) regions as defined in the main text. We have assumed that the single production cross section $\sigma(pp \rightarrow S \rightarrow \gamma\gamma) = 5$~fb.}
\end{figure}

\begin{figure}[!h]
 \centering
\includegraphics[width=0.42\textwidth]{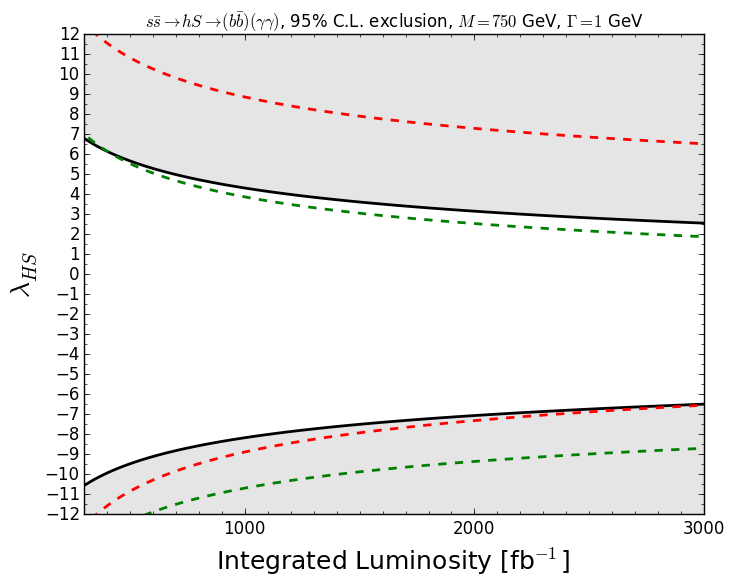}
  \caption{\label{fig:exc_ss_1}  The grey-shaded area shows the 95\% confidence-level exclusion region for the portal coupling $\lambda_{HS}$ for the $s\bar{s}$-induced case and $M=750$~GeV, $\Gamma = 1$~GeV, for the combination of the ``TBD'' (red) and ``OS$\gamma\gamma$'' (green) regions as defined in the main text. We have assumed that the single production cross section $\sigma(pp \rightarrow S \rightarrow \gamma\gamma) = 5$~fb.}
\end{figure}

\begin{figure}[!h]
 \centering
\includegraphics[width=0.42\textwidth]{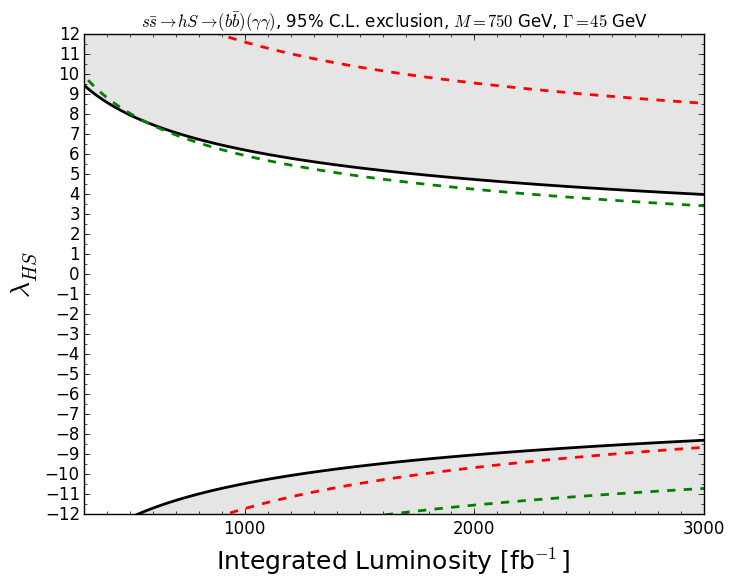}
  \caption{\label{fig:exc_ss_45} The grey-shaded area shows the 95\% confidence-level exclusion region for the portal coupling $\lambda_{HS}$ for the $s\bar{s}$-induced case and $M=750$~GeV, $\Gamma = 45$~GeV, for the combination of the ``TBD'' (red) and ``OS$\gamma\gamma$'' (green) regions as defined in the main text. We have assumed that the single production cross section $\sigma(pp \rightarrow S \rightarrow \gamma\gamma) = 5$~fb.}
\end{figure}

\begin{figure}[!h]
 \centering
  \includegraphics[width=0.42\textwidth]{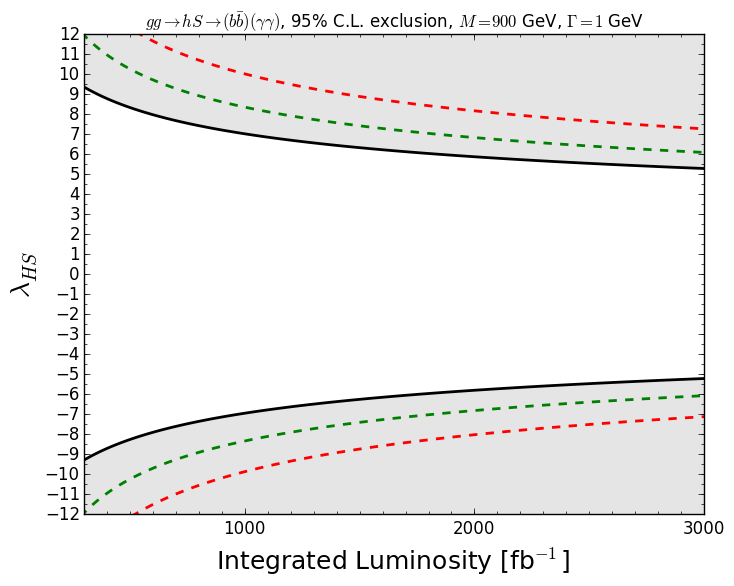}
  \caption{\label{fig:exc_gg_900_1} The grey-shaded area shows the 95\% confidence-level exclusion region for the portal coupling $\lambda_{HS}$ for the $gg$-induced case and $M=900$~GeV, $\Gamma = 1$~GeV, for the combination of the ``TBD'' (red) and ``OS$\gamma\gamma$'' (green) regions as defined in the main text. We have assumed that the single production cross section $\sigma(pp \rightarrow S \rightarrow \gamma\gamma) = 1$~fb.}
\end{figure}

\begin{figure}[!h]
 \centering
  \includegraphics[width=0.42\textwidth]{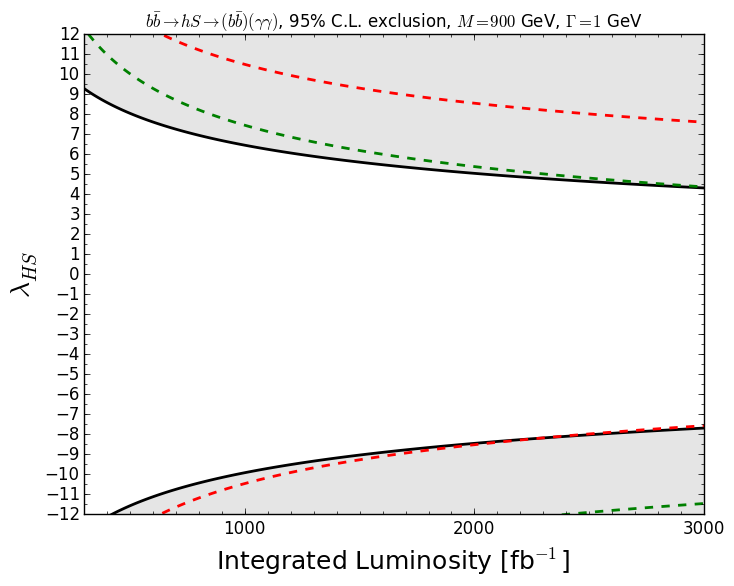}
  \caption{\label{fig:exc_bb_900_1} The grey-shaded area shows the 95\% confidence-level exclusion region for the portal coupling $\lambda_{HS}$ for the $b\bar{b}$-induced case and $M=900$~GeV, $\Gamma = 1$~GeV, for the combination of the ``TBD'' (red) and ``OS$\gamma\gamma$'' (green) regions as defined in the main text. We have assumed that the single production cross section $\sigma(pp \rightarrow S \rightarrow \gamma\gamma) = 1$~fb.}
\end{figure}

\begin{figure}[!h]
 \centering
  \includegraphics[width=0.42\textwidth]{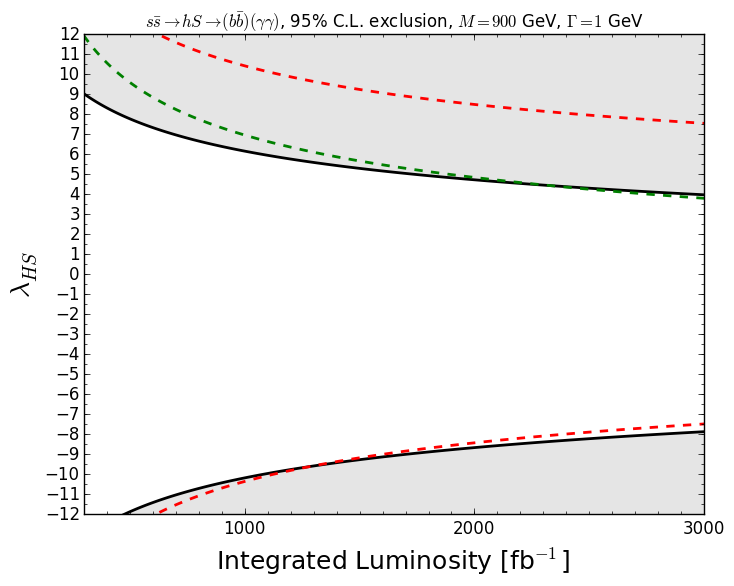}
  \caption{\label{fig:exc_ss_900_1} The grey-shaded area shows the 95\% confidence-level exclusion region for the portal coupling $\lambda_{HS}$ for the $s\bar{s}$-induced case and $M=900$~GeV, $\Gamma = 1$~GeV, for the combination of the ``TBD'' (red) and ``OS$\gamma\gamma$'' (green) regions as defined in the main text. We have assumed that the single production cross section $\sigma(pp \rightarrow S \rightarrow \gamma\gamma) = 1$~fb.}
\end{figure}

We show the resulting expected limits (assuming our null hypothesis is true) as a function of the integrated luminosity for the different
benchmark scenarios that we consider in
Figs.~\ref{fig:exc_gg_600_1}-\ref{fig:exc_ss_900_1}. For the case $M=750$~GeV we show results for $\Gamma = 45$~GeV as well. For $\Gamma = 1$~GeV, we obtain more stringent
constraints, limiting, for $M=600,~750,~900$~GeV respectively, $|\lambda_{HS}| \lesssim 2,~4,~5$ for the
$gg$-initiated process and $\lambda_{HS} \in [\sim  -4, \sim 1],~ [\sim  -7, \sim 3],~[\sim -8, \sim 4]$, both for the $b\bar{b}$ and  $s\bar{s}$-initiated processes, at the end of the HL-LHC run (3000~fb$^{-1}$). The variation between the results for
$b\bar{b}$ and $s\bar{s}$-initiated processes -- visible in the plots -- is very small and can be attributed to the differences
between the parton density functions for the strange and bottom quarks, as already mentioned.

The scenario with the larger width, $M=750$~GeV, $\Gamma = 45$~GeV, clearly exhibits weaker
constraints, with the $gg$-initiated processes yielding
$|\lambda_{HS}| \lesssim 5$, the $b\bar{b}$-initiated process
$\lambda_{HS} \in [\sim -8, \sim 5]$ and the $s\bar{s}$-initiated
process $\lambda_{HS} \in [\sim -8 , \sim 4]$. 
It is conceivable that if further decay channels of the resonance $S$ are
discovered, the remaining unconstrained regions in the $q \bar q$ cases can be covered (in particular 
for narrow width), allowing determination of the
initial state partons that produce the resonance.

The lower bound on $\lambda_{HS}$ for the $q\bar{q}$-initiated cases
is driven by the ``TBD'' region. This is understood by the fact that the ``TBD'' region always vanishes near $\lambda_{HS} \sim 0$, as it is dominated by the diagram with an on-shell $s$-channel $S$, making the exclusion region resulting from it symmetric with respect to $\lambda_{HS} \sim 0$, whereas the ``OS$\gamma\gamma$'' region possesses a symmetry point somewhere in $\lambda_{HS} < 0$.

\subsection*{Mixed production of the di-photon resonance}

So far we have investigated production of the scalar resonance initiated purely either via gluon fusion or $q\bar{q}$ annihilation. We can generalize this to ``mixed'' production via $gg$ and $q\bar{q}$ initial states simultaneously. We concentrate on the scenario of $gg$ and $b\bar{b}$ for simplicity, with the extension to additional quark flavours being straightforward. In that case, the ratio of cross sections, $\rho_\mathrm{mixed}$, defined between the associated production and single production modes can be written as:
\begin{eqnarray}
\rho_\mathrm{mixed} &=& \frac{ \sigma(b\bar{b}\rightarrow hS \rightarrow h \gamma \gamma) +\sigma(gg\rightarrow hS \rightarrow h \gamma \gamma)  } {  \sigma(b\bar{b} \rightarrow S 
\rightarrow \gamma \gamma) + \sigma(gg\rightarrow S \rightarrow \gamma \gamma) } \nonumber \\ 
&=& \frac{B_2 (\lambda_{HS}) [(y^S_d)^{33}]^2 + G_2(\lambda_{HS}) (c_G^S)^2}{B_1 [(y^S_d)^{33}]^2 + G_1 (c_G^S)^2} \;, 
\end{eqnarray}
where $B_{2}(\lambda_{HS})$, $G_2(\lambda_{HS})$ are functions of the portal coupling $\lambda_{HS}$ and $B_1$, $G_1$ are constants with respect to the portal coupling, all to be determined. Defining $\theta \equiv |[(y^S_d)^{33}] / c_G^S|$, the above expression can be re-written as:
\begin{equation}
\rho_\mathrm{mixed} = \frac{B_2 (\lambda_{HS}) \theta^2 + G_2(\lambda_{HS}) }{B_1 \theta^2 + G_1} \;.
\end{equation}
By considering the limits $\theta \rightarrow 0$ and $\theta \rightarrow \infty$, and dividing the numerator and denominator by $G_1$, we can deduce that
\begin{equation}\label{eq:rhomixed}
\rho_\mathrm{mixed} = \frac{ \rho(b\bar{b}, \lambda_{HS})   r_{bg} \theta^2 + \rho(gg, \lambda_{HS})  }{r_{bg} \theta^2 + 1} \;,
\end{equation}
where $\rho(b\bar{b}, \lambda_{HS})$ and $\rho(gg, \lambda_{HS})$ are
the ratios of cross sections as functions of the portal coupling as
defined in Eq.~\ref{eq:rho}, for the cases of pure production
initiated either via $b\bar{b}$ or $gg$, respectively, and $r_{bg}$ is
the ratio of pure single production cross sections for $\theta = 1$:
$r_{bg} = B_1/G_1$. The former two functions have already been
determined in the analysis of the pure cases. The constant $r_{bg}$
was calculated explicitly via Monte Carlo to be $r_{bg} \approx
1.02,~0.69,~0.49$ for $M=600,~750,~900$~GeV respectively. The values
are approximately equal for both the narrow width case ($\Gamma =
1$~GeV) and the larger width case ($\Gamma  = 45$~GeV). 

\begin{figure}[!h]
 \centering
\includegraphics[width=0.42\textwidth]{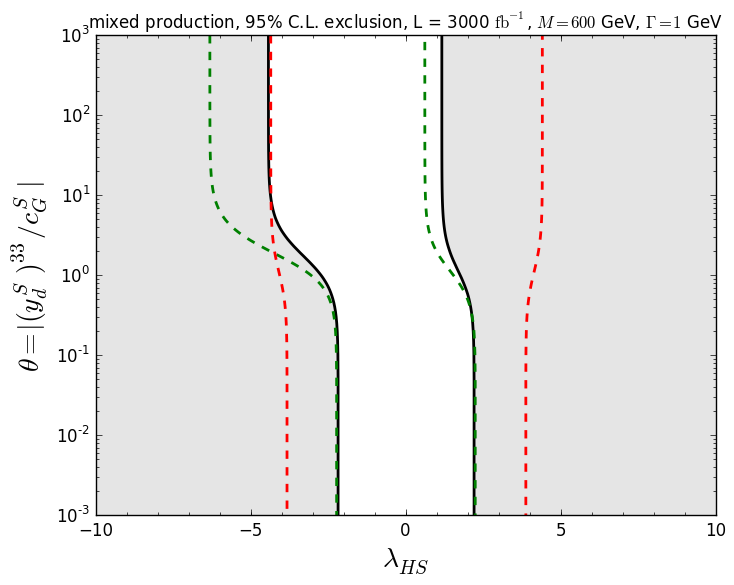}
  \caption{\label{fig:twodim_600} The 95\% confidence-level exclusion region for the ratio of couplings to $b\bar{b}$ over the coupling to the $gg$ initial states, $\theta = [(y^S_d)^{33}] / c_G^S$, versus the portal coupling $\lambda_{HS}$ for $M=600$~GeV, $\Gamma = 1$~GeV. The excluded region coming from the combination of  the ``TBD'' (red) and ``OS$\gamma \gamma$'' (green) regions is grey-shaded. We have assumed that the single production cross section $\sigma(pp \rightarrow S \rightarrow \gamma\gamma) = 10$~fb.}
\end{figure}

\begin{figure}[!h]
 \centering
\includegraphics[width=0.42\textwidth]{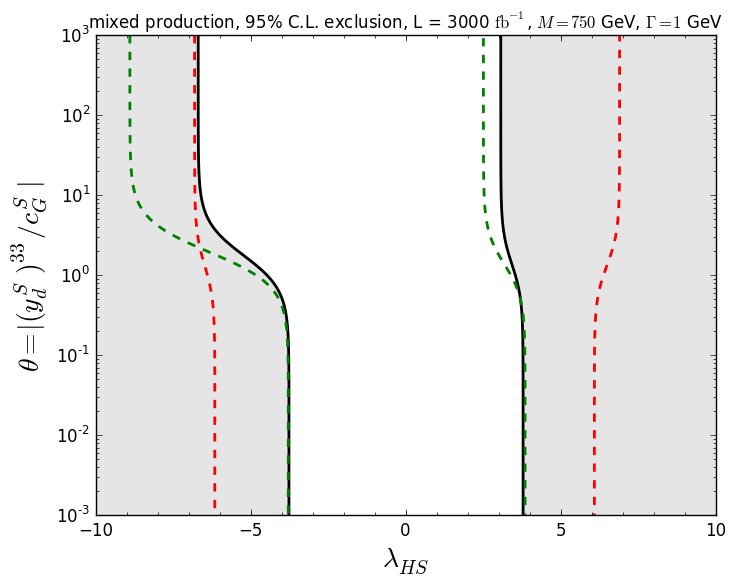}
  \caption{\label{fig:twodim_750} The 95\% confidence-level exclusion region for the ratio of couplings to $b\bar{b}$ over the coupling to the $gg$ initial states, $\theta = [(y^S_d)^{33}] / c_G^S$, versus the portal coupling $\lambda_{HS}$ for $M=750$~GeV, $\Gamma = 1$~GeV. The excluded region coming from the combination of  the ``TBD'' (red) and ``OS$\gamma \gamma$'' (green) regions is grey-shaded. We have assumed that the single production cross section $\sigma(pp \rightarrow S \rightarrow \gamma\gamma) = 5$~fb.}
\end{figure}

\begin{figure}[!h]
 \centering
\includegraphics[width=0.42\textwidth]{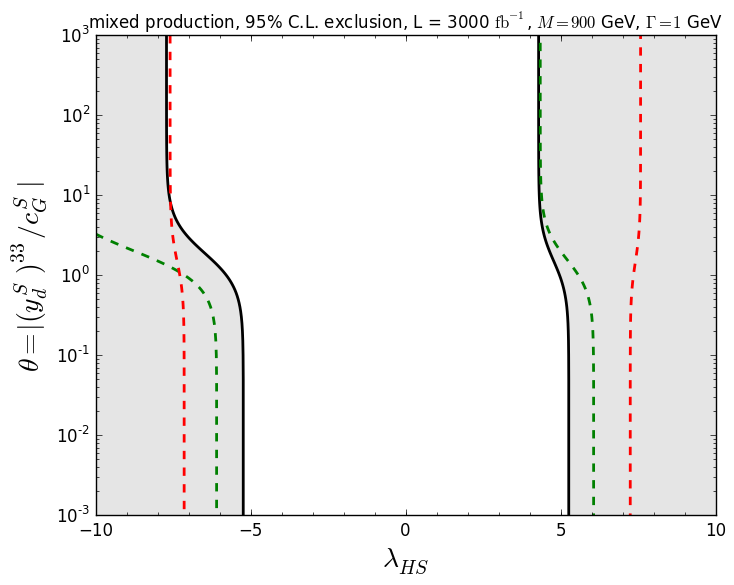}
  \caption{\label{fig:twodim_900} The 95\% confidence-level exclusion region for the ratio of couplings to $b\bar{b}$ over the coupling to the $gg$ initial states, $\theta = [(y^S_d)^{33}] / c_G^S$, versus the portal coupling $\lambda_{HS}$ for $M=900$~GeV, $\Gamma = 1$~GeV. The excluded region coming from the combination of  the ``TBD'' (red) and ``OS$\gamma \gamma$'' (green) regions is grey-shaded. We have assumed that the single production cross section $\sigma(pp \rightarrow S \rightarrow \gamma\gamma) = 1$~fb.}
\end{figure}

\begin{figure}[!h]
 \centering
\includegraphics[width=0.42\textwidth]{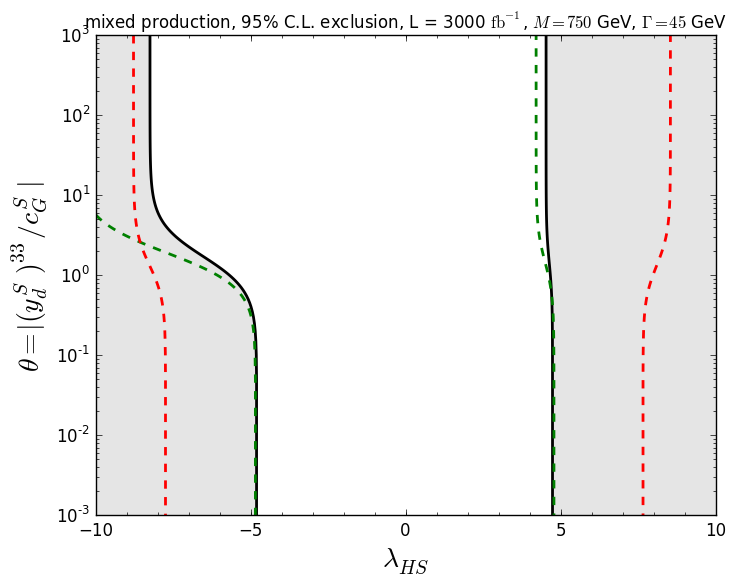}
  \caption{\label{fig:twodim_wide} The 95\% confidence-level exclusion region for the ratio of couplings to $b\bar{b}$ over the coupling to the $gg$ initial states, $\theta = [(y^S_d)^{33}] / c_G^S$, versus the portal coupling $\lambda_{HS}$ for $M=750$~GeV, $\Gamma = 45$~GeV. The excluded region coming from the combination of  the ``TBD'' (red) and ``OS$\gamma \gamma$'' (green) regions is grey-shaded. We have assumed that the single production cross section $\sigma(pp \rightarrow S \rightarrow \gamma\gamma) = 5$~fb.}
\end{figure}

Using Eq.~\ref{eq:rhomixed}, we can deduce an expression for the predicted number of signal events after the application of analysis cuts in the mixed production case:
\begin{equation}\label{eq:nevents}
N_\mathrm{mixed} = \frac{ r_{bg} \theta^2 } { r_{bg} \theta^2 + 1 } N(b\bar{b}, \lambda_{HS}) + \frac{1}{r_{bg} \theta^2 + 1} N(gg,\lambda_{HS}) \;,
\end{equation}
where $N(b\bar{b}, \lambda_{HS})$ and $N(gg,\lambda_{HS})$ are the expected signal events for either pure $b\bar{b}$ or pure $gg$ production for a given value of $\lambda_{HS}$, at a specific integrated luminosity.

The predicted number of background events after a given set of analysis cuts is constant with $\theta$, apart from the associated production of the di-photon resonance with a $b\bar{b}$ pair, which was found to be significant only in the $gg$-initiated scenario. This background scales as $N_\mathrm{assoc.} = N_{assoc.,0}/(r_{bg} \theta^2 + 1)$, where $N_{assoc.,0}$ is the expected number of events for the $b\bar{b}S$ associated production after analysis cuts in the case of pure $gg$-initiated production. 

Using the expression of Eq.~\ref{eq:nevents} and the event numbers for the two analysis regions ``TBD'' and ``OS$\gamma \gamma$'' obtained for the pure production modes, we can derive constraints on the $(\theta, \lambda_{HS})$-plane. These are shown in Figs.~\ref{fig:twodim_600}-\ref{fig:twodim_900} and~\ref{fig:twodim_wide}, for masses $M=600,~750,~900$~GeV and $\Gamma = 1$~GeV and for $M=750$~GeV and $\Gamma = 45$~GeV, respectively, at an integrated luminosity of 3000~fb$^{-1}$ at 13~TeV. It can be seen that in the limits $\theta \rightarrow 0$ and $\theta \gg 1$, corresponding to $gg$- or $b\bar{b}$-dominated production respectively, one can recover the pure production constraints of Figs.~\ref{fig:exc_gg_600_1}-\ref{fig:exc_ss_900_1} obtained at 3000~fb$^{-1}$.
\section{Discussion and Conclusions.}
\label{sec:conc}
We have investigated the associated production of a di-photon
scalar resonance with a Higgs boson and have employed the $pp
\rightarrow hS \rightarrow (b\bar{b}) (\gamma\gamma)$ final state to
obtain constraints on the portal coupling with the SM Higgs boson
$\lambda_{HS}$, at the LHC. We have considered three benchmark scalar masses, $M = 600,~750,~900$~GeV, and we have assumed
that the inclusive single production cross section is $\sigma(pp
\rightarrow S \rightarrow \gamma\gamma) = 10, 5, 1$~fb, respectively, compatible with current experimental constraints. To construct expected
constraints we considered the null hypothesis ({\it i.e.} the supposed
`true' underlying theory) to correspond to gluon-fusion-initiated
production with vanishing portal coupling, $\lambda_{HS} = 0$. We then first analysed
the case of either pure gluon-fusion-induced production or production via quark-anti-quark annihilation. For gluon-fusion production one can impose constraints on the portal coupling at the end of the HL-LHC run 
of $|\lambda_{HS}| \lesssim 2, 4, 5$ for $M=600,~750,~900$~GeV, respectively, and $\Gamma =
1$~GeV, while $|\lambda_{HS}| \lesssim 5$ in the case of large width
($\Gamma = 45$~GeV) and $M=750$~GeV. For quark-anti-quark annihilation, the production
of an on-shell scalar and the Higgs boson is enhanced by a contact
interaction $\sim q\bar{q}hS$, originating from the same $D = 5$
operator that mediates single ${\cal S}$ production. This implies that
the cross section is non-negligible even for vanishing portal coupling
$\lambda_{HS} = 0$. This fact can be exploited to exclude the whole
plane of $\lambda_{HS}$, thus excluding the hypothesised production
via $q\bar{q}$ annihilation. For the case of $b\bar{b}$ we find that a
region inside $\lambda_{HS} \in [\sim -4, \sim 1],~[\sim -7, \sim 3],~[\sim -8, \sim 4]$ for $M=600,~750,~900$~GeV, respectively, could remain
unconstrained at the end of the HL-LHC in the narrow width scenario,
while $\lambda_{HS} \in [\sim -8, \sim 5]$ for large width and $M=750$~GeV. For the
case of $s\bar{s}$ annihilation we find the same unconstrained regions to good approximation
for narrow width, while we obtain $\lambda_{HS} \in [\sim
-8, \sim 4]$ in the case of large width, where $M=750$~GeV. We
have also considered `mixed' production via gluon fusion and $b\bar{b}$
annihilation and derived constraints on the plane of ratio of the corresponding
couplings versus $\lambda_{HS}$ for an integrated luminosity of
3000~fb$^{-1}$.

Note that since a measurement of $\lambda_{HS}$ can straightforwardly 
be translated into a measurement of $\mu_S$, these numbers suggest that
it is possible to exclude, for example, the scenario
where the mass of $S$ stems from EWSB ($\mu_S=0$), which corresponds to
$\lambda_{HS}=M^2/v^2\approx 5.9, 9.3, 13.4$ (still assuming
linear terms in ${\cal S}$ to vanish).\footnote{This is somewhat similar to the case of testing
the presence/size of the $\mu$ term in the SM Higgs potential in Higgs pair production \cite{Goertz:2015dba}.}
Furthermore, if additional decay modes of the resonance $S$ are discovered beyond $\gamma \gamma$, it is
conceivable that a combined analysis in various channels would be able
to exclude all possible values of $\lambda_{HS}$ - for the case of $q \bar q$ production - thus providing an independent method of
determining the production mode. Conversely, the
analysis of the present article demonstrates that if the
production mechanism is constrained via alternative means, it will be possible
to obtain meaningful constraints on the interaction of a new scalar
resonance and the Higgs boson, allowing determination of its relation
to electroweak symmetry breaking.

\acknowledgments
We would like to thank Peter Richardson, Jernej Kamenik, Jos\'e Zurita, Tania Robens and Eleni Vryonidou for useful comments and discussion. AC acknowledges support by  a Marie Sk\l{}odowska-Curie Individual Fellowship of the European Community's Horizon 2020 Framework Programme for Research and Innovation under contract number 659239 (NP4theLHC14). The research of F.G. is supported by a Marie Curie Intra European Fellowship within the 7th European Community Framework Programme (grant no. PIEF-GA-2013-628224). AP acknowledges support by the MCnetITN FP7 Marie Curie Initial Training Network
PITN-GA-2012-315877 and a Marie Curie Intra European Fellowship within
the 7th European Community Framework Programme (grant
no. PIEF-GA-2013-622071). 
\appendix
\section{The full set of operators.}\label{app:operators}
In this appendix, we provide the full set of operators for the SM in the presence of an additional dynamical scalar singlet ${\cal S}$, up to $D=5$. 
The fact that ${\cal S}$ does not transform under the SM gauge group
makes the construction straightforward: in the case that it is CP even (with CP-conserving interactions), each additional operator will consist of a gauge invariant SM operator, multiplied by powers of ${\cal S}$ (and potentially derivatives).\footnote{We neglect the only $D=5$ operator consisting only of SM building blocks  -- the Weinberg operator -- since the tiny neutrino masses suggest that it is suppressed by a very large scale $\Lambda \sim M_{\rm GUT}$.} This means that in particular that, schematically, ${\cal L} \supset {\cal L}_{\rm SM} + {\cal L}_{\rm SM} \cdot {\cal S}$. If all operators in the SM would be marginal, {\it i.e.}, feature $D=4$, this would already correspond to the full Lagrangian at $D\leq5$, connecting the SM with the new resonance. In turn, the only operators that are missing (up to pure NP terms) are those containing the single relevant ({\it i.e.}, $D<4$) operator of the SM, $|H|^2$. This can be multiplied by up to three powers of ${\cal S}$ as well as by $\partial^2 {\cal S}$ (the latter being equivalent via integration by parts to ${\cal S} \partial^2 |H|^2$). This finally leads to the Lagrangian \footnote{See also~\cite{Gripaios:2016xuo, Franceschini:2016gxv}. We do not include leptons, as they do not enter the examined processes to the order considered. Moreover, note that (induced) tadpoles in ${\cal S}$ can always be removed via a field redefinition.}
\bee
\label{eq:FullL}
\mathcal{L}_{\rm eff}& = & {\cal L}_{\rm SM} +
\frac{1}{2}\partial_\mu {\cal S} \partial^\mu {\cal S} - \frac 1 2 \mu_S^2 {\cal S}^2 \nonumber \\
&-& \frac{\lambda^\prime_{S}}{2 \sqrt 2} v {\cal{S}}^3 - \frac{\lambda_{S}}{4} {\cal{S}}^4 - \lambda^\prime_{HS} v |H|^2 {\cal{S}} - \lambda_{HS} |H|^2 {\cal{S}}^2 \nonumber \\
&-& \frac{\cal{S}}{\Lambda} \left[ c_{\lambda S} {\cal{S}}^4 
+c_{HS} |H|^2 {\cal{S}}^2 
+c_{\lambda H} |H|^4 \right]\nonumber \\ 
&-&(y_d^S)^{ij} \frac{\cal{S}}{\Lambda}\bar{Q}_L^{i}H d_R^{j} - (y_u^S)^{ij}\frac{\cal{S}}{\Lambda}\bar{Q}_L^i\tilde{H}u_R^j +\mathrm{h.c.}\nonumber\\
&-&\frac{\cal{S}}{\Lambda}\frac{1}{16\pi^2}\left[g^{\prime 2} c_B^S B_{\mu\nu} B^{\mu\nu}+ g^2 c_W^S W^{I\mu\nu} W_{\mu\nu}^I\right.\nonumber\\
&+&\left.g_S^2 c_G^S G^{a\mu\nu}G_{\mu\nu}^a\right]\,.
\label{eq:aa1}
\eee

Note that we have used equations of motion/field redefinitions to eliminate the operators $\mathcal{S}/ \Lambda (\partial^\mu {\cal S})^2,\, \mathcal{S}/\Lambda |D_{\mu}H|^2,\, \mathcal{S}/\Lambda  \partial^2|H|^2 $ and those of type ${\cal S}/\Lambda\, \bar{q} D_\mu \gamma^\mu q$.
The couplings in the third line of (\ref{eq:aa1}) and $\lambda^\prime_{S,HS}$, which are those that have been neglected in (\ref{eq:lagrangian}), do not contribute to the process under consideration to good approximation: They either do not enter at leading order, or (in the case of the $|H|^{2,4} {\cal S}$ interactions) at most
contribute to a diagram with a (strongly suppressed) off-shell
Higgs boson propagator, discussed in Appendix \ref{app:fullfit}. The same holds
true for the $S^4$ term, that was kept for the discussion in Section
\ref{sec:intro}.

\section{Fits for the ratio $\rho$ including a linear term.}\label{app:fullfit}

In addition to the diagrams of Figs.~\ref{fig:ggdiags}
and~\ref{fig:qqdiags}, there can be contributions to the final state under consideration
originating from  $hhS$ interactions (for example, an intermediate $s$-channel Higgs boson to
$hS$ or an $S \to hh$ 
intermediate state), provided the
following triple scalar interaction, linear in $S$, is non-vanishing:
\bee
\label{eq:Le}
 - \kappa_{HS} v h^2 S \subset  - \lambda^\prime_{HS} v |H|^2 {\cal{S}}
- c_{\lambda H} \frac{\cal{S}}{\Lambda} |H|^4\;,
\label{eq:hmix}
\eee 
with $\kappa_{HS} = 1/2\, \lambda^\prime_{HS} + 3/(2\Lambda)\,  c_{\lambda H}$.

In this scenario, terms will appear in the cross section ratio proportional to
$\kappa_{HS}$ and $\kappa_{HS}^2$. Due to interference with the new $hhS$ diagrams (note that also the `non-signal' $S \to hh$ interferes with the signal diagrams after an off-shell $h \to \gamma \gamma$), some dependence on the coupling of the resonance $S$ to photons, $c_{\gamma}^S$, as well as the production couplings, $(y_d^S)^{22}$, $(y_d^S)^{33}$ and $c_G^S$, is introduced in the denominator of the cross section ratio. Due to this, smaller values of $|c_{\gamma}^S|$ and the production couplings enhance the effect of the new contributions proportional to $\kappa_{HS}$ and $\kappa_{HS}^2$. To take these effects into account and at the same time remain conservative, we set  $|c_{\gamma}^S|$ to the  possible minimal value that produces a $\sigma(pp\to S\to \gamma\gamma)\sim 5\,\fb$ for $M=750$~GeV, as derived in Ref.~\cite{Gupta:2015zzs}. This leads to  $|c_{\gamma}^S|\gtrsim \{60,\, 100,\, \, 10\}$ for $s\bar{s}$, $b\bar{b}$ and $gg$ production, respectively, where here and in the following we assume a normalisation of $\Lambda=1\,$TeV. Moreover, one can derive conservative lower bounds on $|(y_{d}^{S})^{22}|$, $|(y_d^S)^{33}|$ and $|c_{g}^S|$, by demanding $\sigma(pp \to S)\gtrsim 5\,$fb, leading to $|(y_{d}^{S})^{22}| \gtrsim 0.15$, $|(y_{d}^{S})^{33}| \gtrsim 0.2$ and $|c_g^S|\gtrsim 0.25$. We set these to their minimal values as well in what follows. 

We can then parametrize the ratio 
$\rho(xx')=\sigma(xx' \rightarrow h \gamma
  \gamma)/  \sigma(xx'\rightarrow S 
\rightarrow \gamma \gamma),\,xx'=b\bar b,s \bar s,gg$, by\footnote{For the case of vanishing $\kappa_{HS}$, this more general definition coincides with Eq.~\ref{eq:rho} to good approximation.}
\begin{eqnarray}
\rho(xx') &=& \delta_1^x + \delta_2^x\, \kappa_{HS} \lambda_{HS}  + \delta_3^x\, \kappa_{HS}   \nonumber \\
&+& \delta_4^x\, \lambda_{HS}  +  \delta_5^x\, \kappa_{HS}^2   + \delta_6^x\,  \lambda_{HS}^2 \;,
\end{eqnarray}
where $\delta_i^{b,s,g}$ are coefficients to be determined. 

\begin{table}[!h]
\begin{ruledtabular}
\begin{tabular}{cccc}
$i$ & $\delta_i^b$ & $\delta_i^s$ & $\delta_i^g$ \\\hline
1 &  $8.27\times 10^{-3}$ & $1.05\times 10^{-2}$ & $< \mathcal{O}(10^{-5})$\\ 
2 & $1.93\times 10^{-4}$ & $-2.30\times 10^{-4}$ & $\mathcal{O}(10^{-7})$\\ 
3 & $8.26\times 10^{-4}$  &  $2.54 \times 10^{-5}$& $< \mathcal{O}(10^{-6})$  \\ 
4 &  $3.08\times 10^{-3}$  & $3.71 \times 10^{-3}$ & $< \mathcal{O}(10^{-6})$  \\ 
5 &   $2.40 \times 10^{-5}$ & $6.00 \times 10^{-5}$ & $< \mathcal{O}(10^{-8})$\\ 
6 & $8.60\times 10^{-4}$  & $9.27 \times 10^{-4}$  & $1.33 \times 10^{-3}$   \\ 
\end{tabular}
\caption{\label{tb:fitcs} The $\rho$ fit coefficients for the general case of including an interaction term proportional to $\mathcal{L} \supset -\kappa_{HS} v h^2 S$ for width $\Gamma =1$~GeV and $M = 750$~GeV.}
\end{ruledtabular}
\end{table}

An estimate of the coefficients in this conservative scenario,
obtained numerically, is given in Table~\ref{tb:fitcs} for width $\Gamma =1$~GeV and $M = 750$~GeV. 
The contributions from diagrams with an off-shell Higgs boson are a priori
not fully negligible for the $b\bar{b}$-induced case, simply because of interference
of the $s$-channel $h \to h S$ with the large matrix elements with the on-shell $S$ in the final
state. 

To investigate the size of $\kappa_{HS}$ that renders the
off-shell Higgs boson contributions significant to the $pp \rightarrow
h \gamma \gamma$ process, we consider the ratio $\rho(\kappa_{HS}, \lambda_{HS}) /
\rho(\kappa_{HS}=0, \lambda_{HS}) = 1+r$, where $r$ is a number that
characterises the fractional change in the ratio $\rho$ due to
$\kappa_{HS}$ for a given value of $\lambda_{HS}$. If we choose $r=1$,
which implies $\mathcal{O}(1)$ changes due to $\kappa_{HS}$, and solve
for $\kappa_{HS}$ for values $\lambda_{HS} \sim \mathcal{O}(1)$ we obtain: $|\kappa_{HS} |\sim \{ \mathcal{O}(10), \, \mathcal{O}(10), \,  \mathcal{O}(10^2) \}$ for $s\bar{s}$, $b\bar{b}$ and $gg$ production, respectively.

Since the gauge-invariant terms in the Lagrangian generating this
coupling also induce a Higgs-scalar mixing, they can not be arbitrarily
large. Indeed, Higgs data can constrain $|\kappa_{HS}|\lesssim 4$ at 95\%
confidence-level \cite{Carmi:2012in, Cheung:2015dta,
  Bauer:2016lbe}.\footnote{Here, we have neglected the impact of the second operator in (\ref{eq:Le}), ${\cal S}|H|^4$, which breaks the correlation between Higgs-scalar mixing and the $h^2 S$ interaction.} Therefore, given the values calculated in the previous paragraph, for $\kappa_{HS}$ to eventually have a non-negligible effect on the analysis of the present article, this bound needs to be violated, and setting $\kappa_{HS} = 0$ is a justified approximation. Despite the fact that we focussed on $M=750$~GeV, the analysis is expected to give similar estimates for the other benchmark points considered in the main analysis of this article, $M=600,~900$~GeV.


\newpage 

\bibliography{diphotonres}
\bibliographystyle{apsrev4-1}

\end{document}